\documentclass{aastex}
\usepackage{epsfig}
\def\lsim{\raise0.3ex\hbox{$<$}\kern-0.75em{\lower0.65ex\hbox{$\sim$}}}
\def\gsim{\raise0.3ex\hbox{$>$}\kern-0.75em{\lower0.65ex\hbox{$\sim$}}}

\begin{document}

\title{
Evidence For Cloud-Cloud Collision and Parsec-Scale Stellar
Feedback Within the L1641-N Region
}
\author{Fumitaka Nakamura\altaffilmark{1,2}, 
Tomoya Miura\altaffilmark{3}, Yoshimi Kitamura\altaffilmark{2}, 
Yoshito Shimajiri\altaffilmark{4}, Ryohei Kawabe\altaffilmark{4}, 
Norio Ikeda\altaffilmark{2}, Takashi Tsukagoshi\altaffilmark{5}, 
Munetake Momose\altaffilmark{6}, Ryoichi Nishi\altaffilmark{3}, and
Zhi-Yun Li\altaffilmark{7}
}
\altaffiltext{1}{National Astronomical Observatory, Mitaka, Tokyo 181-8588, 
Japan; fumitaka.nakamura@nao.ac.jp}
\altaffiltext{2}{Institute of Space and Astronautical Science, 
Japan Aerospace Exploration Agency, 3-1-1 Yoshinodai, Sagamihara, 
Kanagawa 229-8510, Japan}
\altaffiltext{3}{Department of Physics, 
Niigata University, 8050 Ikarashi-2, Niigata, 950-2181, Japan}
\altaffiltext{4}{Nobeyama Radio Observatory, Minamimaki, Minamisaku, 
Nagano 384-1805, Japan}
\altaffiltext{5}{Institute of Astronomy, Faculty of Science, 
University of Tokyo, Osawa 2-21-1, Mitaka, Tokyo, 181-0015, Japan}
\altaffiltext{6}{Institute of Astrophysics and Planetary Sciences, 
Ibaraki University, Bunkyo 2-1-1, Mito 310-8512, Japan}
\altaffiltext{7}{Department of Astronomy, University of Virginia,
P. O. Box 400325, Charlottesville, VA 22904; zl4h@virginia.edu}
\begin{abstract}
We present high spatial resolution $^{12}$CO ($J=1-0$) images 
taken by the Nobeyama 45m telescope 
toward a $48' \times 48'$ area including the L1641-N cluster. 
The effective spatial resolution of the maps is 
$21''$, corresponding to 0.04 pc at a distance of 400 pc.
A recent 1.1 mm dust continuum map reveals that 
the dense gas is concentrated in several thin filaments.
We find that a few dust filaments are located at the 
parts where $^{12}$CO ($J=1-0$) emission drops sharply. 
Furthermore, the filaments have two-components with different 
velocities.  
The velocity difference between the two-components is 
about 3 km s$^{-1}$, corresponding to a Mach number of 10, 
significantly larger than the local turbulent velocity in the cloud. 
These facts imply that the collision of the two components (hereafter,
the cloud-cloud collision) possibly contributed to the formation of 
these filaments.
Since the two components appear to overlap toward the filaments 
on the plane of the sky,
the collision may have occurred almost along the line of sight.
Star formation in the L1641-N cluster
was probably triggered by such a collision.
We also find several parsec-scale CO shells whose centers
are close to either the L1641-N cluster or 
V 380 Ori cluster. We propose that these shells were 
created by multiple winds and/or outflows from cluster YSOs, 
i.e., ``protocluster winds.'' 
One exceptional dust filament located at 
the western cloud edge lies along a shell; 
it is presumably a part of the expanding shell.
Both the cloud-cloud collision and protocluster winds are likely to 
influence the cloud structure and kinematics in this region.
\end{abstract}

\section{Introduction}

\label{sec:intro}

Most stars form in giant molecular clouds (GMCs).
In GMCs, various environmental effects such as 
large-scale flows, supernovae, and stellar feedback 
from young stars (winds, radiation, and outflows) 
often shape the cloud structure and dynamics, 
triggering and suppressing the formation of the next-generation
stars \citep[e.g.,][]{mckee07}.  
Recent numerical simulations of star formation have 
demonstrated that these effects significantly influence,  
or even control, the cloud evolution and star formation  
\citep{maclow04, krumholz11,li06,nakamura07,banerjee09,
carroll09,gritschneder09,enrique10}.
Fingerprints of these environmental effects have been 
found in some star-forming regions
\citep[e.g.,][]{bally89,heyer92,bally08,odell08,sandell01, shimajiri08, arce10, shimajiri10}. 
However, the roles of the environmental effects in star formation 
remain poorly understood observationally.
This is partly because wide-field, high spatial and/or spectral 
resolution observations, which are needed to resolve the cloud 
structure and kinematics in detail, are still limited. 
In particular, stellar feedbacks such as winds or outflows are 
often extended to parsec-scale \citep[e.g.,][]{heyer92}. Wide-field 
observations of the cloud structure and kinematics are needed to unveil
such environmental effects. At the same time, it is necessary to 
resolve the cloud structure at a scale of ``dense cores'',
which are the basic units of individual star formation ($\approx $ 0.1 pc 
$\sim $ one arcmin at a distance of 400 pc),
in order to uncover a link between individual star formation and 
the environmental effects.

In this paper, to understand how the environmental effects influence
the internal structure and kinematics in star-forming molecular clouds, 
we present the results of wide-field $^{12}$CO ($J=1-0$) mapping observations
toward the L1641-N region, a nearby active star-forming region 
in the Orion A giant molecular cloud complex, 
using the Nobeyama 45 m radio telescope. 
Our data have high angular ($\approx 21''$) resolution, allowing us to resolve
spatial structures at a scale of 0.04 pc at a distance of 400 pc
\citep[see e.g.,][]{menten07,sandstrom07,hirota08}.  
By comparing the $^{12}$CO ($J=1-0$) map with the 1.1 mm dust 
continuum map taken by the AzTEC camera on the ASTE telescope, 
we investigate how the dense gas is distributed 
in the parent molecular cloud.

The L1641-N region is one of the well-studied, nearby star-forming regions
\citep[][and references therein]{allen08}. 
It lies just south the clouds known as OMC-4 and OMC-5, 
making up a northern part of the L1641 molecular cloud, 
one of the several $\sim 10^4M_\odot$ molecular clouds 
contained within the Orion A giant molecular cloud complex.
The L1641 cloud is a filamentary cloud extending 
more than 2 $-$ 3 degrees ($15\sim 20$ pc) and having numerous clumpy 
elongated condensations with typical masses of a few tens 
to $\sim 10^2M_\odot$ and sizes of a few tenths to $\sim$ 1 pc.
This cloud is not forming massive stars: its most massive member
is the B4 V Herbig Ae/Be star, HD 38023,  
at the position of (R.A., Dec.) = (5:42:21, $-$8:08), 
located at the L1641-S region.
Although it has no rich clusters comparable to 
the Orion Nebula Cluster, the L1641-N region consists of 
several active star-forming regions. Optical, X-ray, 
and infrared surveys have revealed that most of the protostars 
are clustered in small clusters or groups with a few $\times 10$ YSOs
\citep{strom93,carpenter01}. 
A prominent example is the L1641-N cluster which 
contains approximately 80 YSOs \citep{galfalk08,fang09}, 
having the stellar surface density of $\sim 200$ pc$^{-2}$,
an order of magnitude larger than the average stellar surface density 
of the distributed population of YSOs.
The most luminous object in this cluster is IRAS 05338-0624, around which 
powerful molecular outflows were discovered \citep{fukui86,stanke07}.

Another example is the V 380 Ori cluster. 
V 380 Ori itself is a binary of Herbig Ae/Be stars, illuminating 
the reflection nebula NGC1999 \citep{baines06}. 
Many Harbig-Haro (H-H) objects were discovered in this region
\citep[][and references therein]{allen08}. 
The most famous and spectacular outflows are
H-H 1/2, which are associated with gigantic bow shocks.  
\citet{heyer92} discovered a parsec-scale expanding shell around 
the V 380 Ori star from the $^{13}$CO ($J=1-0$) observations,
suggesting that star formation activity indeed shapes cloud 
structure and dynamics significantly.
These two clusters do not contain massive stars
that would emit strong UV radiation and control the evolution of the region.
The L1641-N region is expected to provide us with clues
to understanding the impact of the current and previous star formation 
activity on the cloud structure and kinematics.

The rest of the paper is organized as follows.
The details of our observations and data are described in 
Sections \ref{sec:obs} and \ref{sec:data}.
We present results of our $^{12}$CO $(J=1-0)$ observations 
in Section \ref{sec:results}, and discuss the cloud structure and 
star formation activity of this region in Section \ref{sec:discussion}.
Finally, we summarize our conclusion in Section \ref{sec:summary}.

\section{Observations}
\label{sec:obs}

The $^{12}$CO ($J=1-0; 115.271204$ GHz) observations were carried out 
with the 25-element focal plane receiver BEARS on the
Nobeyama Radio Observatory (NRO) 45 m telescope. 
They cover a $48'\times 48'$ area 
including the L1641-N cluster, 
the northern part of the L1641 molecular cloud,  
from 2009 December to 2010 January. 
At 115 GHz, the telescope has a FWHM beam size of 15$''$ 
and a main beam efficiency, $\eta$, of 0.32. 
At the back end, we used 25 sets of 1024 channel auto-correlators 
(ACs) which have bandwidths of 32 MHz and frequency resolutions 
of 37.8 kHz. The frequency resolution corresponds 
to a velocity resolution of 0.1 km s$^{-1}$ at 115 GHz. 
During the observations, the system noise temperatures were 
in the range between 300 to 600 K in DSB at the observed elevations. 
The standard chopper wheel method was used to convert the output signal 
into the antenna temperatures ($T_{\rm A}^*$), corrected for the atmospheric 
attenuation.  Our mapping observations were made by the 
On-The-Fly (OTF) mapping technique. We adopted a spheroidal 
function as a gridding convolution function (GCF) to calculate the intensity
at each grid point of the final cube data with a spatial grid size of
7.5$''$ and a velocity resolution of 0.5 km s$^{-1}$.  
The final effective resolution of the map is 21$''$,
corresponding to 0.04 pc 
at the distance to Orion A of 400 pc.
The rms noise level of the final map is 0.39 K in $T_{\rm A}^*$.

\section{Other Data}
\label{sec:data}

In the next section, we compare our $^{12}$CO ($J=1-0$) data 
with 1.1 mm continuum, $^{13}$CO ($J=1-0$), and 
H$^{13}$CO$^+$ ($J=1-0$) data.
Here, we briefly describe these data.

The 1.1 mm continuum data were taken toward a $1.7^\circ \times
2.3^\circ$ region in the northern part of the Orion A giant molecular cloud 
complex with the AzTEC camera mounted on 
the ASTE 10 m telescope \citep{shimajiri10}. 
The observations were carried out in the period 
from October to December 2008.
The noise level was about 9 mJy beam$^{-1}$ and the 
effective beam size was 40$''$, about twice that of the $^{12}$CO map, 
after the FRUIT imaging
which is an iterative mapping method to recover the 
spatially-extended component.
The detail of the data is given in \citet{shimajiri10}.

The $^{13}$CO ($J=1-0$) data were taken by \citet{bally87}
with the 7 m telescope of the AT\&T Bell Laboratories.
The original data covers the whole Orion A giant molecular cloud complex
\citep[see][]{bally87,bally89}, much larger area than
that of our $^{12}$CO ($J=1-0$) data.
The observations were carried out in the period from
1984 to 1986.
The grid size of the data was 60$''$, about three times coarser 
than that of our $^{12}$CO ($J=1-0$) data.
The rms noise level was about 0.3 K  in $T_A^*$  
at the velocity resolution of about 0.27 km s$^{-1}$.
The main beam efficiency of the 7 m telescope was 0.9.

The  H$^{13}$CO$^+$ ($J=1-0$) data were taken from the Nobeyama 45 m 
archival data. The data were obtained 
from December 1999 to April 2004 in the position-switching mode, 
with a grid size of 21$''$.
The rms noise level was 0.1 K in $T_A^*$ 
at the velocity resolution of about 0.13 km s$^{-1}$
with a main-beam efficiency of 0.51.
See http://www.nro.nao.ac.jp/ for more detail.

\section{Results}
\label{sec:results}

\subsection{Global $^{12}$CO ($J=1-0$) and $^{13}$CO ($J=1-0$) Distributions}

\begin{figure}[h]
\epsscale{0.45}
\plotone{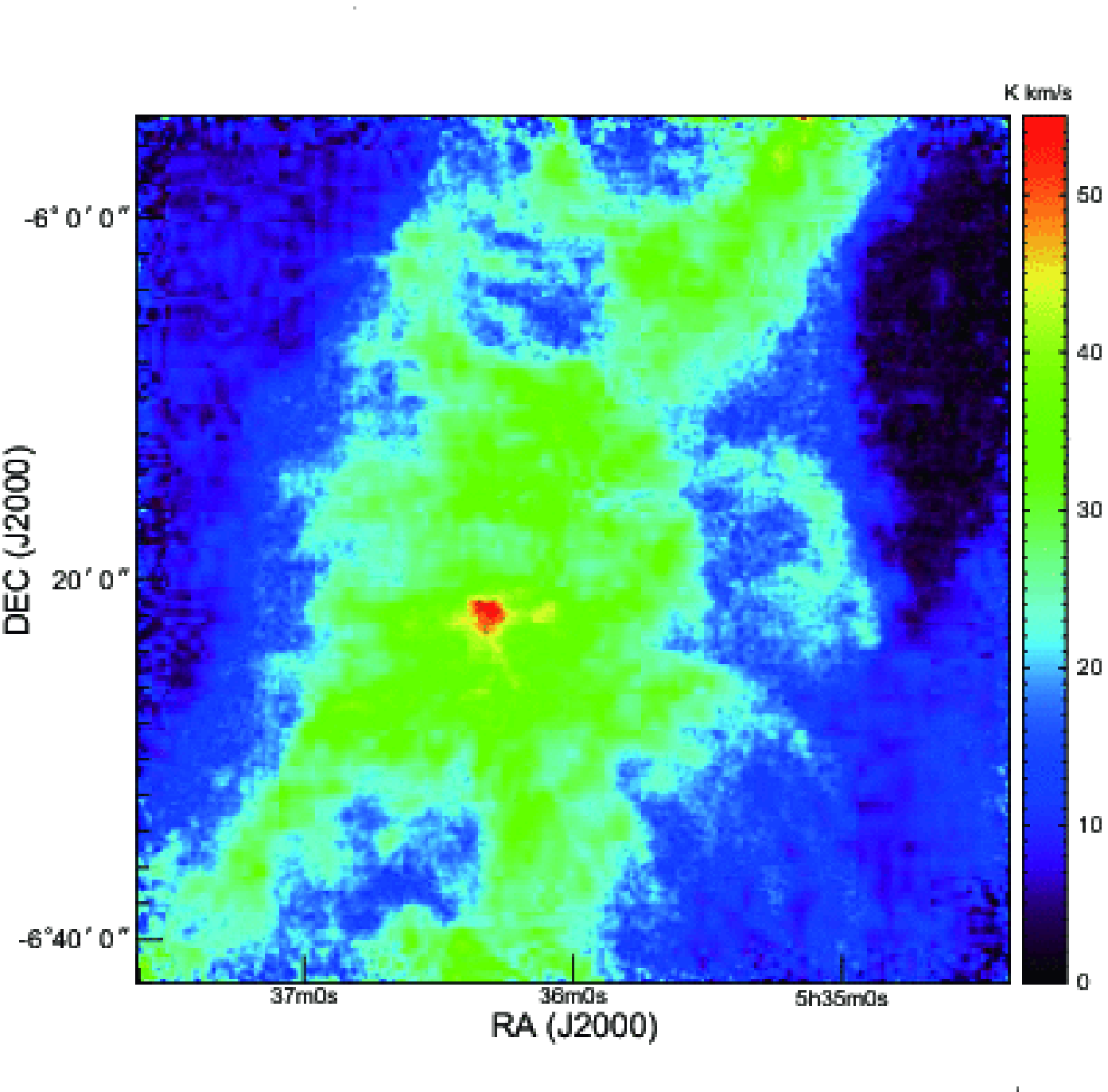}
\plotone{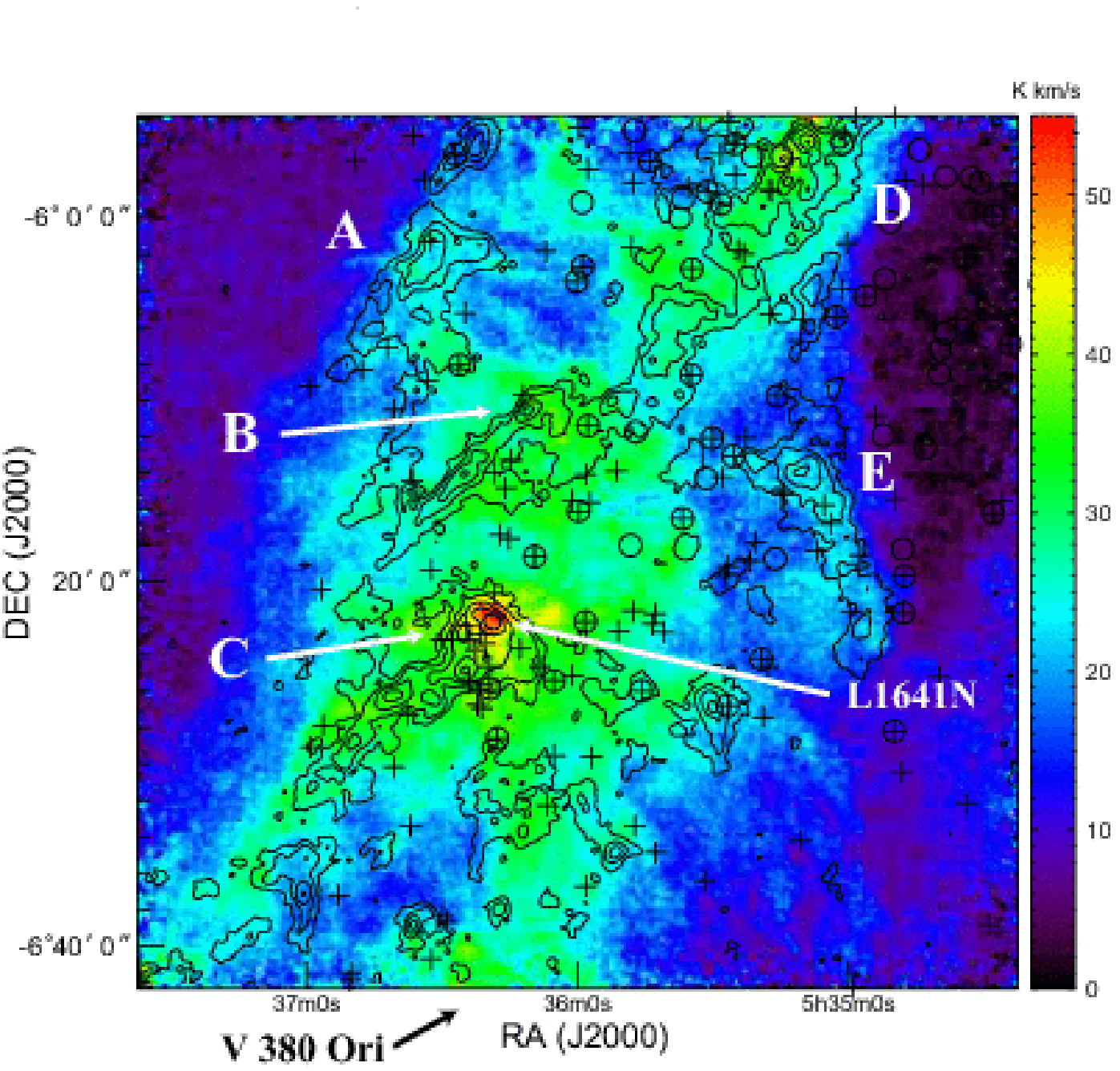}
\caption{(a) $^{12}$CO ($J=1-0$) total integrated intensity map in 
the velocity range from $v_{\rm LSR}=$ 0.0 to 20.0 km s$^{-1}$
toward the $48'\times 48'$ area including the L1641-N cluster. 
The effective spatial resolution of the map is $21''$.
The $^{12}$CO emission peak coincides reasonably well 
with the position of the L1641-N cluster.
(b) Same as panel (a), but the 1.1 mm continuum data 
are overlaid on the $^{12}$CO ($J=1-0$) total integrated intensity map. 
The contours are drawn at 0.06, 0.15, 0.3, 0.6, and 1.2 Jy/beam.  
The 1 $\sigma$ rms noise level is 9 mJy/beam.
The positions of YSOs identified by \citet{rebull06} and 
\citet{carpenter01} are overlaid on the images 
by the circles and crosses, respectively.
Note that the plotted YSOs are less embedded sources, and
therefore younger embedded YSOs are not shown.
Several dust filaments are labeled with alphabet A through E. 
The positions of L1641-N and V 380 Ori are indicated by the white arrows.
}  
\label{fig:map1}
\end{figure}

In Figure \ref{fig:map1}, we present a $^{12}$CO ($J=1-0$) total 
integrated intensity map toward the $48' \times 48'$ area 
including the L1641-N cluster. 
The $^{12}$CO ($J=1-0$) emission tends to be optically thick in the 
entire molecular cloud, thus representing the spatial extent
of the global molecular gas distribution, instead of 
the density distribution.
For comparison, in Figure \ref{fig:map1}(b), the 1.1 mm continuum map 
taken by the AzTEC camera mounted on the ASTE 10 m telescope is 
overlaid on the $^{12}$CO ($J=1-0$) total integrated intensity map
with the contours.

\begin{figure}[h]
\epsscale{0.5}
\plotone{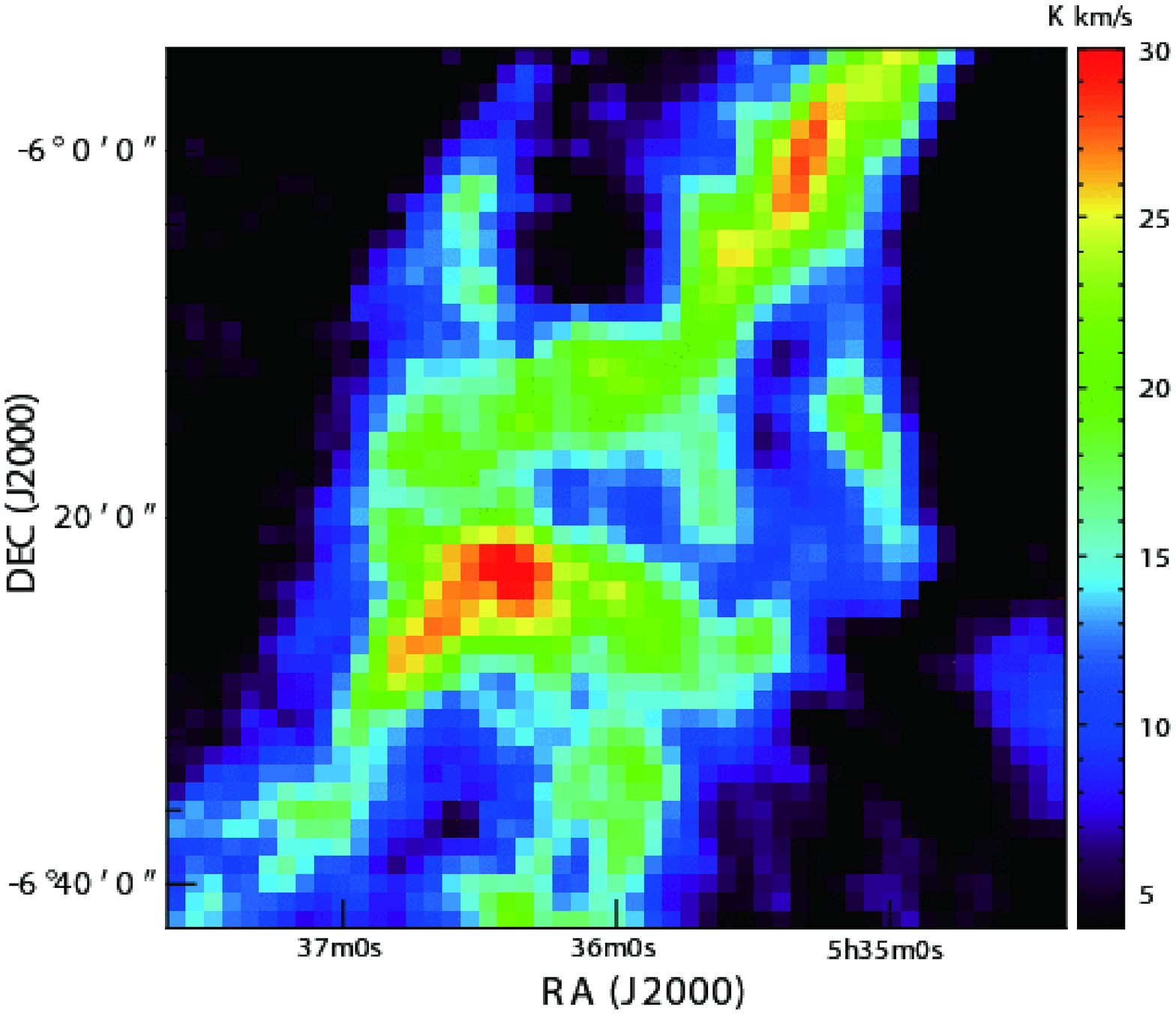}
\plotone{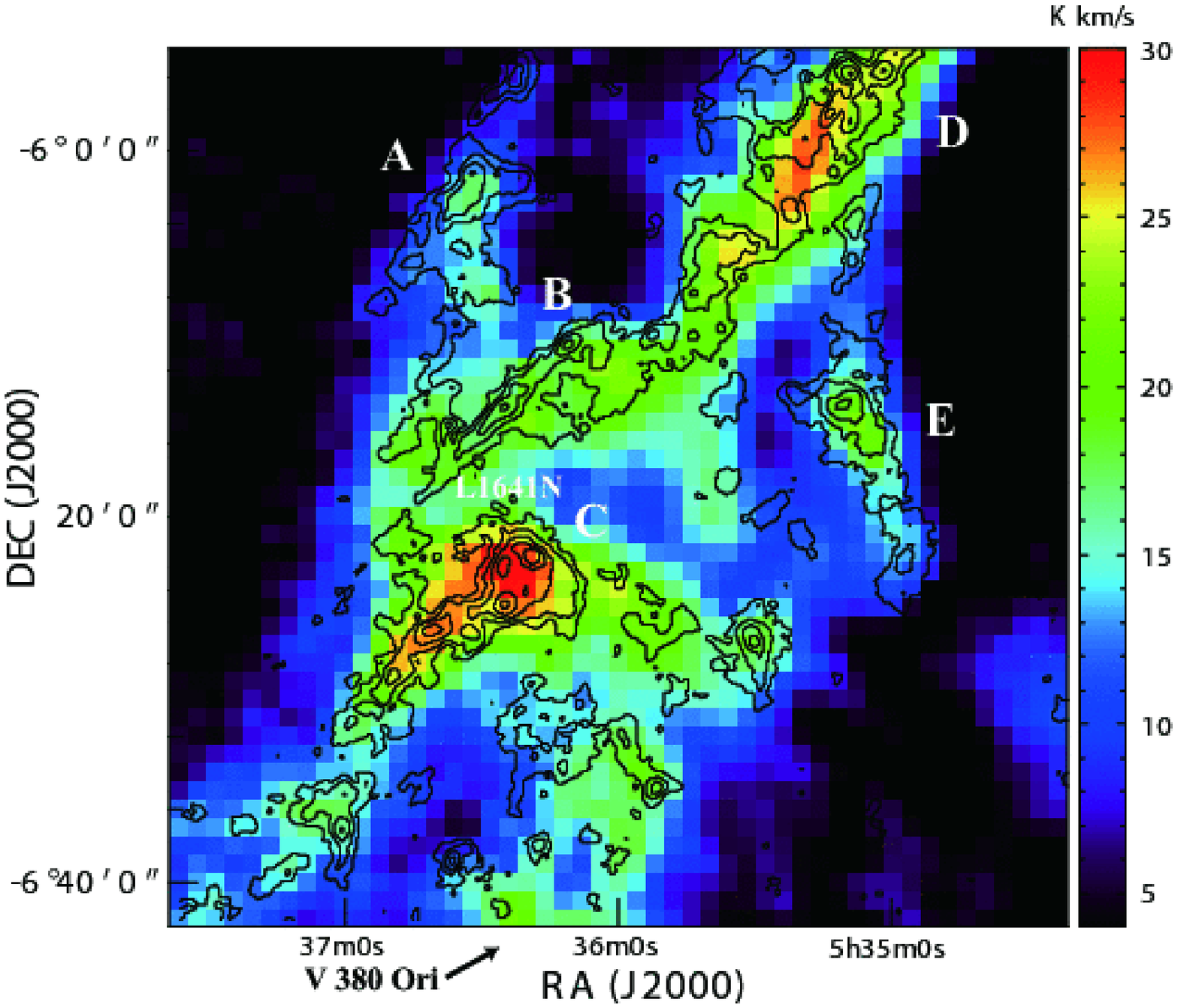}
\caption{(a) $^{13}$CO ($J=1-0$) total integrated intensity map
toward the same area presented in Figure \ref{fig:map1}
in the velocity range from $v_{\rm LSR}=$ 1.0 to 14.0 km s$^{-1}$.
The $^{13}$CO data were obtained by \citet{bally87} with
the 7 m AT\&T Bell laboratories telescope.
The strongest $^{13}$CO emission is associated with the L1641-N cluster.
(b) Same as panel (a) but the 1.1 mm continuum data 
are overlaid with the contours 
on the $^{13}$CO ($J=1-0$) total integrated intensity map. 
The contours are drawn at 0.06, 0.15, 0.3, 0.6, and 1.2 Jy/beam.
Several dust filaments are labeled with alphabet A through E.
}  
\label{fig:13co map}
\end{figure}

The $^{12}$CO ($J=1-0$) integrated intensity map indicates that 
the parent filamentary cloud is roughly running from north to south.
The filament appears to bifurcate in the southern part of the image.
At the bifurcation point, the L1641-N cluster resides.
The most intense $^{12}$CO emission is associated with the L1641-N cluster
at the position of (R.A., Dec.) = (5:36:18, $-$6:21:48).
The position of the $^{12}$CO integrated intensity peak almost 
coincides with that of 
the bright, compact continuum source, L1641-N MM1 identified by
\citet{stanke07} with the SMA interferometer, 
although it deviates about 22$''$ toward south.
This is probably because the single-dish continuum peak
traces the whole clumpy structures identified by SMA, 
including L1641-N MM1, as shown in Figure 6 of \citet{stanke07}.
Comparison between the 1.1 mm continuum image and the $^{12}$CO integrated
intensity map indicates that several filamentary dense clumps
are located near the parts where the $^{12}$CO emission drops sharply.
These parts are in good agreement with the edges traced by $^{13}$CO ($J=1-0$) emission
(see Fig. \ref{fig:13co map}) and therefore, hereafter, we 
refer to these parts as the cloud edges, 
for simplicity.  
For comparison, several main filamentary dense clumps are labeled 
with alphabet A through E in Figure \ref{fig:map1}(b), indicating that 
the filaments A and E are located at the eastern and western cloud
edges, respectively.

In Figure \ref{fig:13co map}, 
we also present the $^{13}$CO ($J=1-0$) integrated intensity map toward
the same area shown in Figure \ref{fig:map1}. 
Since the fractional abundance of $^{13}$CO is smaller than
that of $^{12}$CO by a factor of about 40, the $^{13}$CO emission 
tends to be optically thinner than the $^{12}$CO emission, 
and thus its spatial distribution is expected to represent the distribution 
of molecular gas with intermediate densities of about $10^3$ cm$^{-3}$.
Although the spatial extent of the $^{13}$CO integrated intensity is well
covered by that of $^{12}$CO, its spatial distribution
appears different from that of $^{12}$CO.  The $^{13}$CO emission 
tends to be concentrated in elongated or filamentary structures,
which reasonably trace the parsec-scale filamentary clumps 
identified by the 1.1 mm dust continuum emission 
(see Figure \ref{fig:13co map}).
The 1.1 mm dust continuum emission and $^{13}$CO ($J=1-0$) integrated intensity 
also take their maxima nearly at the same position, 
i.e., toward the L1641-N cluster center, 
suggesting that the dense gas is concentrated into the L1641-N
cluster-forming filamentary clump (the filament C).

\begin{figure}[h]
\epsscale{0.5}
\plotone{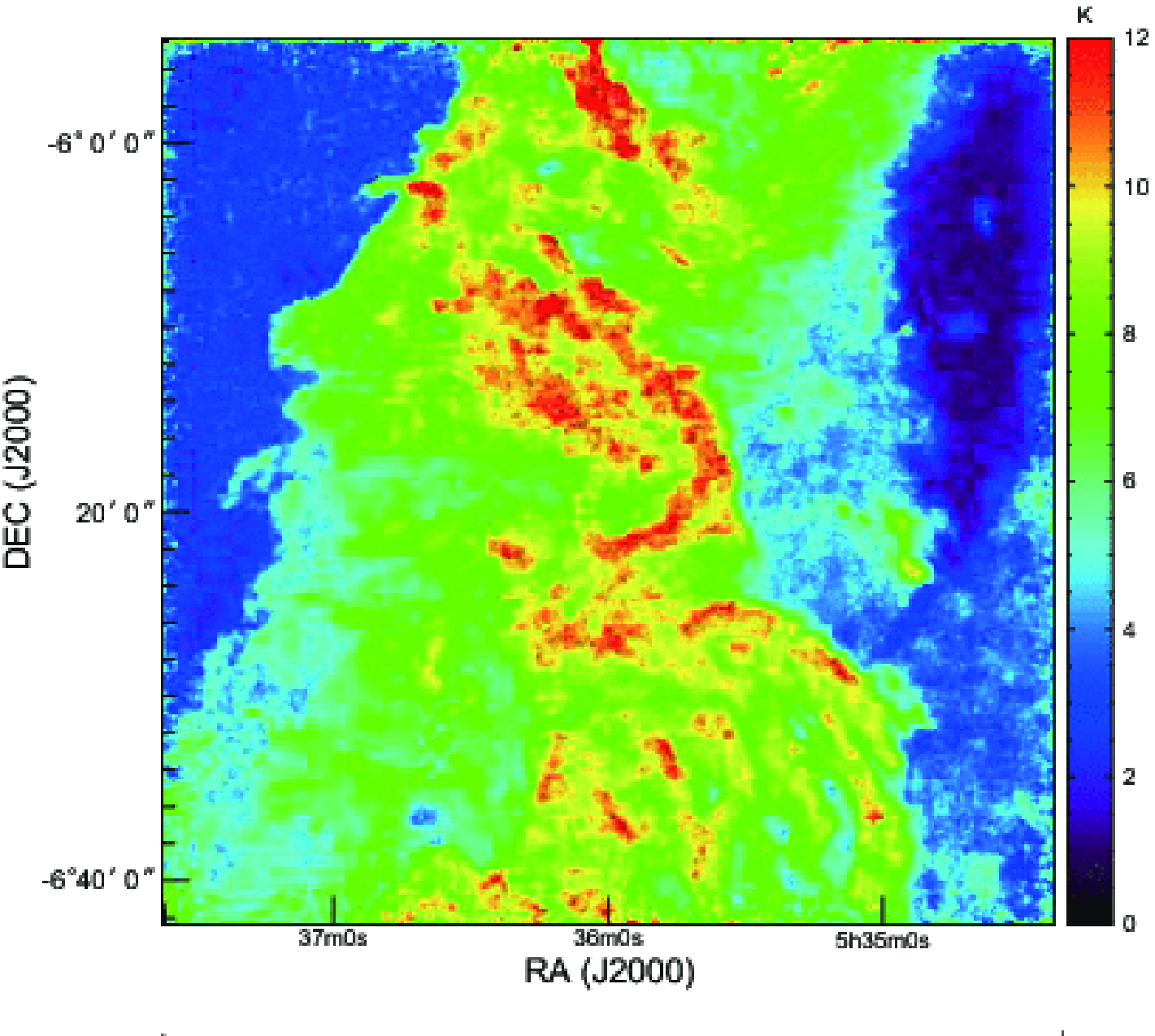}
\plotone{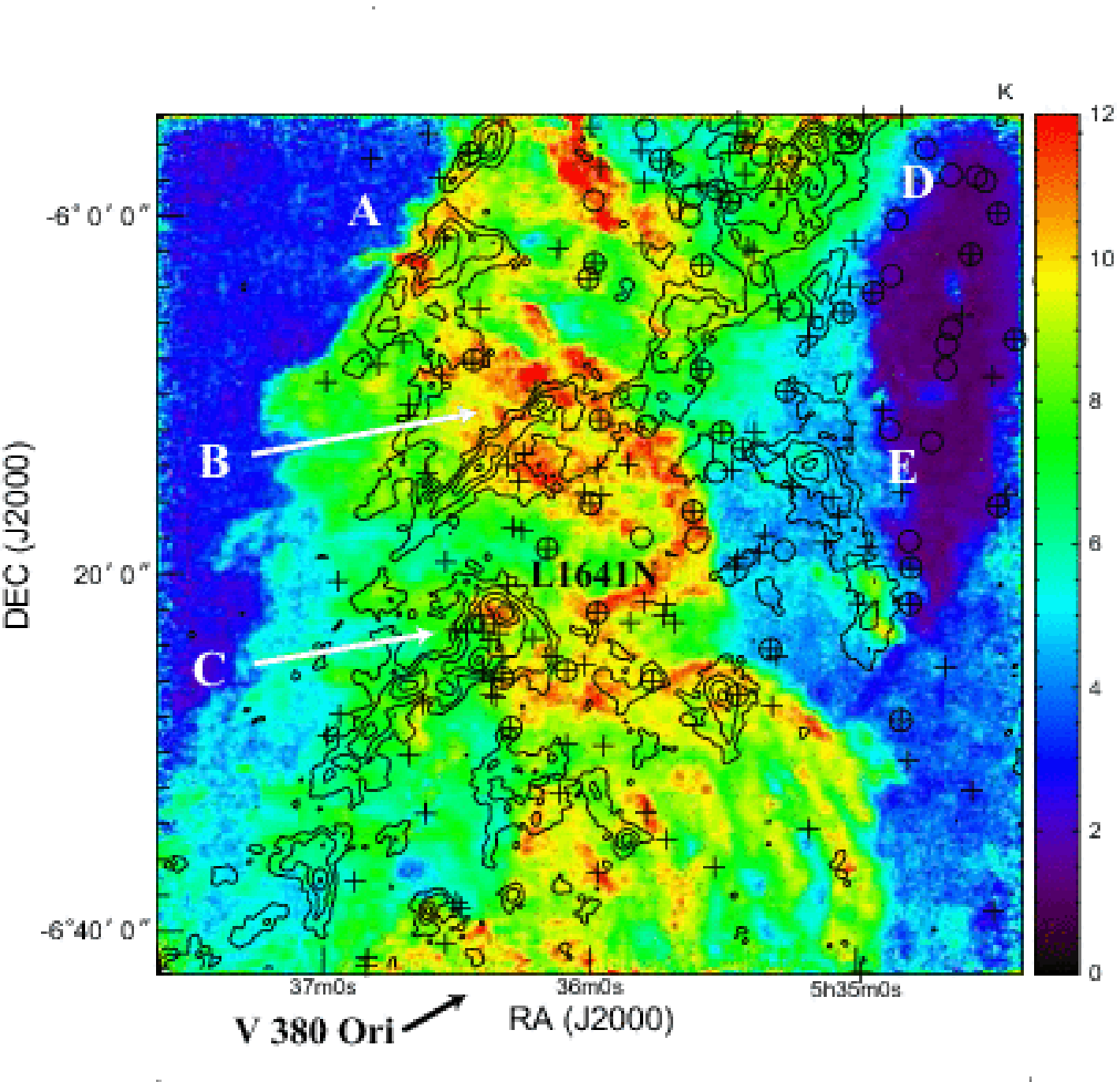}
\caption{(a) $^{12}$CO ($J=1-0$) peak intensity map.
(b) Same as panel (a) but the 1.1 mm continuum contours are overlaid.
The contours are drawn at 0.06, 0.15, 0.3, 0.6, and 1.2 Jy/beam.
The crosses and circles defnotes the positions of YSOs identified by
Spitzer \citep{rebull06}, respectively.
Note that these YSOs are less embedded sources.
Several dust filaments are labeled with alphabet A through E. 
}  
\label{fig:peak map}
\end{figure}

In Figure \ref{fig:peak map}, we present a $^{12}$CO ($J=1-0$) peak 
intensity map, which indicates that the $^{12}$CO ($J=1-0$) emission
drops sharply both at the eastern and western sides of the
 parent filamentary cloud. At the same places, the $^{12}$CO 
and $^{13}$CO integrated intensity drops sharply.
A prominent feature is the existence of many arcs or elongated
structures, most of which tend to be across the filamentary
structures traced by the $^{13}$CO ($J=1-0$) integrated intensity map.
The peak intensity map of the optical thick $^{12}$CO ($J=1-0$) emission
typically represents the distribution of the excitation temperature, 
instead of the local density distribution.
However, the structures traced by the $^{12}$CO ($J=1-0$) peak 
intensity map are sometimes seen in the $^{13}$CO ($J=1-0$) integrated
intensity map and is therefore likely to reflect the cloud density
distribution in some parts.
For example, some arcs located at the southwest part near the cloud edge 
are recognized in both the $^{12}$CO ($J=1-0$) peak intensity and
$^{13}$CO integrated intensity maps 
(e.g., (R.A., Dec.) $\simeq$ (5:35:0, $-$6:40:0) and (5:35:30,$-$6:35:30)).

\subsection{Dust Filaments}
\label{subsec:filament}

Here, we present several aspects of the parsec-scale 
filamentary dust clumps, 
by comparing the spatial distributions of several
molecular emission lines such as $^{12}$CO ($J=1-0$) 
and $^{13}$CO ($J=1-0$) with the 1.1 mm continuum emission.

\begin{figure}[h]
\epsscale{0.8}
\plotone{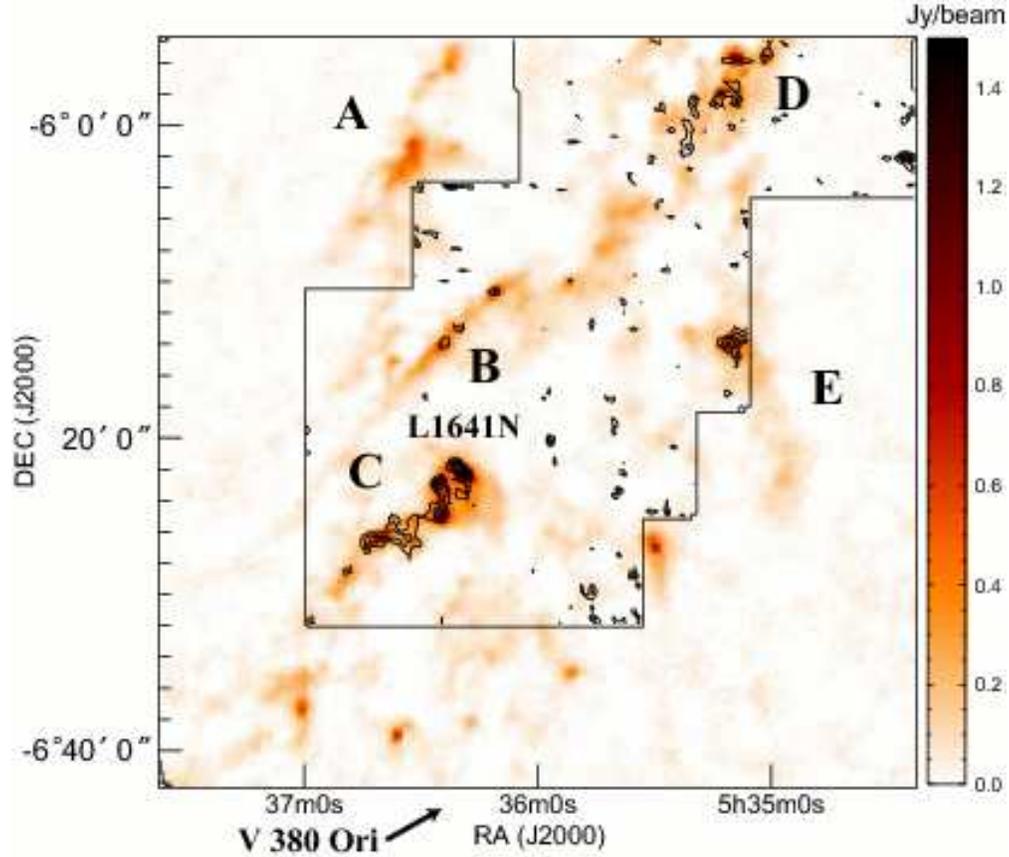}
\caption{H$^{13}$CO$^+$ ($J=1-0$) velocity integrated intensity 
contours overlaid on the 1.1 mm dust continuum image taken by AzTEC on ASTE.
The velocity range is from $v_{\rm LSR}=$ 4.0 to 14.0 km s$^{-1}$.
The contours start from 0.42 K km s$^{-1}$ at intervals of 0.2 K km
 s$^{-1}$. The solid lines indicate the observation box for 
the H$^{13}$CO$^+$ ($J=1-0$) emission.  The data were taken from
the Nobeyama 45 m  archival data (http://www.nro.nao.ac.jp/).
Several dust filaments are detected in the H$^{13}$CO$^+$ emission,
indicating that they contain dense gas with $10^{5}$ cm$^{-3}$.
Several dust filaments are labeled by alphabet A through E.
}  
\label{fig:dust}
\end{figure}

The dust filaments are likely to have high densities 
$\sim 10^{4-5}$ cm$^{-3}$ because the CS ($J=1-0$) emission
is detected toward all the dust filaments 
(see Figure 5(c) of Tatematsu et al. 1993).  
The H$^{13}$CO$^+$ ($J=1-0$) emission is also detected 
toward the dense parts of the filaments \citep{ikeda07}. 
For comparison, we show in Figure \ref{fig:dust}
the H$^{13}$CO$^+$ ($J=1-0$) integrated intensity contours
overlaid on the 1.1 mm image, in which
the dust filaments are labeled with A through E. 
Figure \ref{fig:dust} indicates that the strongest H$^{13}$CO$^+$
($J=1-0$) emission comes from the most massive filament C associated 
with the L1641-N cluster.  The H$^{13}$CO$^+$ emission peaks
in this filament coincide reasonably well with those of the 
1.1 mm continuum emission.  The line-of-sight velocities of the 
H$^{13}$CO$^+$ ($J=1-0$) peaks are around 7 km s$^{-1}$.
Interestingly, the CS ($J=1-0$) emission is strongest 
at $V_{\rm LSR} \sim $ 10 km s$^{-1}$, but
weaker at $V_{\rm LSR} \sim $ 7 km s$^{-1}$ (see Figure 5(c) of Tatematsu et
al. 1993), although the H$^{13}$CO$^+$ emission is strong at around 7
km s$^{-1}$.
In addition, the CS emission associated with the massive filament C
is distributed in the wide velocity range of 
6 km s$^{-1}$ to 12 km s$^{-1}$.
As mentioned below, the CS emission as well as $^{12}$CO indicates that 
the most massive filament C has two different velocity components,
which may give us a clue to understand the formation mechanism of 
the filament.
We note that the filament A is out of the area observed 
by \citet{ikeda07}, but it is detected in CS ($J=1-0$) obtained by 
\citet{tatematsu93}. Thus, the filament contains the
dense gas with densities of $10^{4-5}$ cm$^{-3}$.

\begin{figure}[h]
\epsscale{1.0}
\plotone{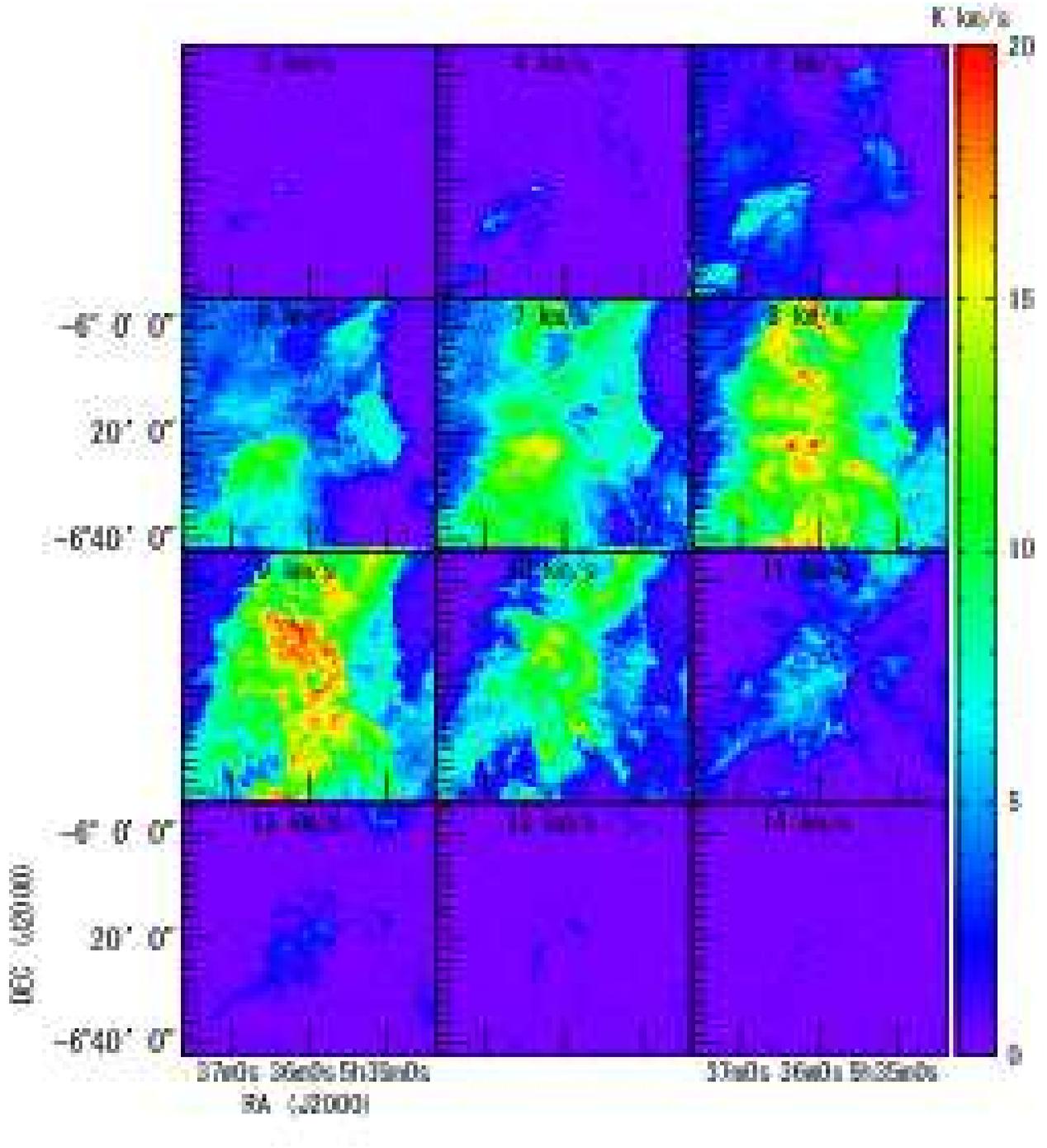}
\caption{$^{12}$CO ($J=1-0$) velocity channel maps with 
velocity width of 1.0 km s$^{-1}$. 
}  
\label{fig:channel}
\end{figure}

\begin{figure}[h]
\epsscale{1.0}
\plotone{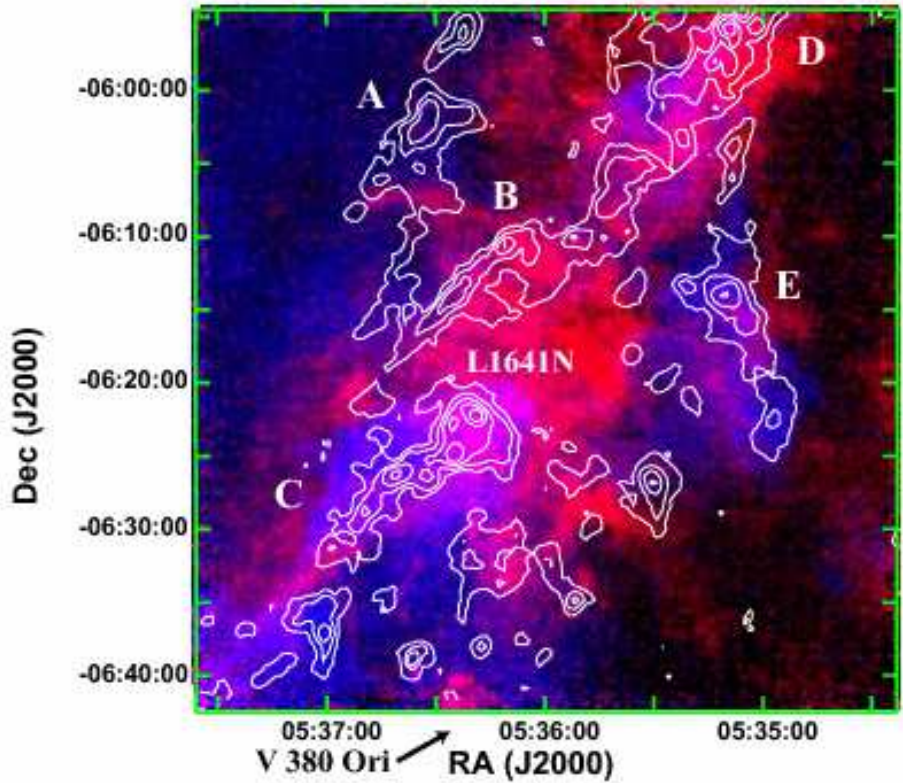}
\caption{Two-color image toward the same area presented in 
Figure \ref{fig:map1} with redshifted $^{12}$CO ($J=1-0$)
integrated intensity in red (9.5 km s$^{-1}$ $\sim 14.5$ km s$^{-1}$) 
and blueshifted $^{12}$CO ($J=1-0$) integrated intensity in blue
(3.5 km s$^{-1}$ $\sim 6.5$ km s$^{-1}$).
The white contours indicate the 1.1 mm dust continuum emission
at levels of 0.06, 0.15, 0.3, 0.6, and 1.2 Jy/beam.
}  
\label{fig:3color}
\end{figure}

The $^{12}$CO ($J=1-0$) velocity channel maps presented in 
Figure \ref{fig:channel} indicate that the velocity structure 
in this region can be divided into 
three components: blueshifted (3$-$6 km s$^{-1}$), 
main (7$-$9 km s$^{-1}$), and redshifted ones (10$-$14 km s$^{-1}$).
For comparison, we present in Figure \ref{fig:3color} 
a two-color image with redshifted $^{12}$CO integrated intensity in red and  
blueshifted $^{12}$CO integrated intensity in blue.
The 1.1 mm continuum map is overlaid on the image with white contours.
The integration ranges are 9.5 $\sim$ 14.5 km s$^{-1}$ and 
3.5 $\sim$ 6.5 km s$^{-1}$ for the redshifted and blueshifted components, 
respectively.
The two-color image indicates that the redshifted component is
dominant in the upper half of the parent filament, whereas
the blueshifted component is dominant in the lower half.

\begin{figure}[h]
\epsscale{0.7}
\plotone{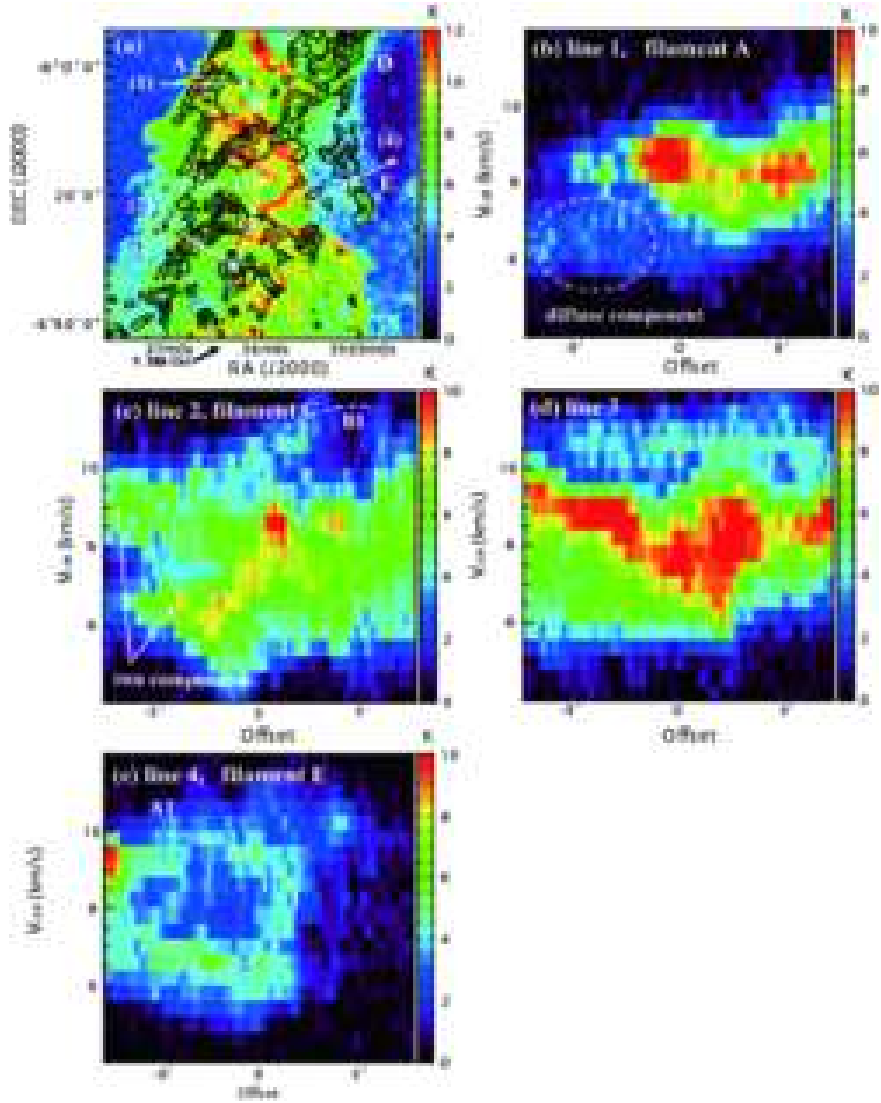}
\caption{(a) $^{12}$CO $(J=1-0)$ peak intensity map showing the
 positions of the PV diagrams presented in panels (b) through (e). 
The 1.1 mm continuum map is overlaid on the image with the 
black contours whose levels are the same as those of Figure \ref{fig:map1}.
(b) Position-velocity diagram along the line (1). 
The abscissa denotes the offset measured from the position of 
the white cross indicated in panel (a).
The plus and minus values are for the western and eastern sides 
of the cross.
(c) Same as panel (b) but for the line (2). 
(d) Same as panel (b) but for the line (3).
(e) Same as panel (b) but for the line (4).}  
\label{fig:PV}
\end{figure}

The strong blueshifted component with a velocity range between 4 and 6 km
s$^{-1}$ appears in the southern part where the filament C and 
the L1641-N cluster reside. The redshifted component with a 
velocity of 10 to 12 km s$^{-1}$ is also associated with the 
massive filament C.  
The presence of a couple of velocity components toward the filament C 
is more clearly seen in the position-velocity map 
presented in Figure \ref{fig:PV}(c), showing that the filament consists of 
two separate components with different velocities toward 
the eastern part of the filament:
one is about 6 km s$^{-1}$ and the other is 9 km s$^{-1}$.
These two-components with different velocities are in good agreement
with the CS map by \citet{tatematsu93}.
The filament is surrounded by the blueshifted elongated 
component that apparently have a head-tail shape 
(see also Figure \ref{fig:3color}).
Another faint velocity component is seen toward the western part of the
filament at the velocity of about 11 km s$^{-1}$. This component 
is more evident in the position-velocity diagrams of the 
$^{13}$CO ($J=1-0$) emission 
(see Figure  \ref{fig:PV2}(c)).

Other dust filaments also have two-component velocity structures.
The position-velocity diagram (Figure \ref{fig:PV}(b)) indicates 
that the filament A, 
located at the eastern cloud edge, has two-components with 
different velocities: a diffuse component is at 6 km s$^{-1}$ 
and stronger emission component is at 9 km s$^{-1}$. 
The similar velocity components are associated with the 
filaments B and D (see also Figure \ref{fig:channel}).  
The two-color image indicates that the eastern area outside
the parent cloud is filled with the diffuse component with 
about 6 km s$^{-1}$.  According to \citet{sakamoto97} and 
\citet{shimajiri10},
this diffuse component is distributed in the much larger area 
in the eastern area of the Orion A giant molecular cloud complex.
The filament E has the two-components at 7 km s$^{-1}$ and 9.5 km s $^{-1}$
(see Figure \ref{fig:PV}(e)). These velocity structures of the filaments
are in good agreement with the position-velocity map of the 
$^{13}$CO ($J=1-0$) emission (see Figure \ref{fig:PV2}(e)).
It is worth noting that the two filaments labeled with
A and E are located near the cloud edges where the $^{12}$CO
($J=1-0$) emission drops sharply. 

\subsection{Shells}

Besides the filamentary structures, shell-like structures 
are prominent in our $^{12}$CO map, particularly in the  
peak intensity map.  
In this subsection, we describe some characteristics of 
these shell-like structures in detail.

In the $^{12}$CO ($J=1-0$) integrated intensity map
(Figure \ref{fig:map1}),
a shell-like structure can be recognized  at the position of 
(R.A., Dec.) $\sim$ (5:36:9, $-$6:4:20). 
The radius of the shell is about 10$'$, corresponding 
to about 1 pc at a distance of 400 pc.
The shell is more clearly seen in the $^{13}$CO ($J=1-0$) integrated 
intensity map presented in Figure \ref{fig:13co map}.
This shell was first found by \citet{heyer92} who carried out a 
wide-field mapping observation in $^{13}$CO ($J=1-0$) toward 
the Orion A giant molecular cloud complex with the FCRAO 14 m telescope.
They found that the shell has two velocity components with a 
velocity difference of about 2$-$3 km s$^{-1}$, implying 
that it has an expanding motion.
They also found a number of large holes surrounded by expanding shells. 
The typical radius of the shells is $10'-26'$, corresponding to 1$-$3 pc
at a distance of 400 pc. They interpreted that these structures
stem from events associated with the energetic star formation 
activity within the cloud.

Similar shell-like structures can be seen in our $^{12}$CO maps. 
For example, in the central part of the $^{12}$CO peak intensity map, 
two prominent thin shell structures are seen (Figure \ref{fig:peak map}).
We labeled these shells by A1 and A2 on the peak intensity map presented
in Figure \ref{fig:peak map2}. 
There is also a shell labeled by A3, which is almost parallel 
to the A1 and A2 shells.
These three shells are spatially well-ordered and 
appear to have a common center which is very close to the position 
of the L1641-N cluster. 
These shells have very small thickness of about 30$''$. 
Such a thin shell may be difficult to be found in the $^{13}$CO map
because of the spatial resolution.

Another prominent shell-like structure can be recognized  
in the southern part of the CO integrated intensity maps
(Figures \ref{fig:map1} and \ref{fig:13co map}), where 
the CO emission appears weak.
Here, we labeled three remarkable shell-like structures or arcs 
by B1, B2, and B3, 
although there are several similar arcs seen in the image 
(Figure \ref{fig:peak map2}).
These arcs appear to be spatially well-ordered and homocentric.
Some of them are detected in the CS ($J=1-0$) line
\citep[see Figure 5(c) of ][]{tatematsu93}, indicating that 
the dense gas is associated with them.
We note that the common center of these shells is very close to
the position of the V 380 Ori cluster, which 
is just outside our $^{12}$CO map, 
located at (R.A., Dec.) = (5:36:25.43, $-$6:42:57.7).

The shells are also clearly seen in 
a volume rendering image of the $^{12}$CO ($J=1-0$) antenna temperature data
presented in Figure \ref{fig:3d}.
The volume rendering technique is suitable to find and visualize
the coherent structures in the 3D space.
The green color typically represents
the areas with relatively high $T_{\rm A}^*$ of about 10$-$15 K.
We designated the shells A1, A2, A3, B1, B2, and B3 in the image.
These shells are coherent in the position-position-velocity cube,
suggesting that they were created by some dynamical events.
Several arcs, parts of the shells, can also be recognized in the 
$^{13}$CO map presented in Figure \ref{fig:13co map}.
These coherent structures are remarkable since
the surrounding gas is highly turbulent and  
such coherent structures would have been destroyed in a crossing time
of the shells (about $10^4$ yr for the radius of $10''-30''$ and
velocity of 3 km s$^{-1}$).
Their existence suggests that the events producing these coherent 
structures play a dominant role in determining the cloud 
internal structure.
In Section \ref{subsec:shell} we will discuss these structures in detail.

\subsection{Molecular Outflows in the L1641N cluster}

Recent numerical simulations of cluster formation have 
suggested that protostellar outflow feedback plays an important
role in regulating star formation in cluster-forming clumps
because they inject substantial amount of kinetic energy into
the surroundings \citep{li06,nakamura07,carroll09,wang10}.
Here, we attempt to identify molecular outflows associated with the 
L1641-N cluster to assess the roles of the outflow feedback 
in the dynamical evolution
of the cloud.
The $^{12}$CO emission has been successfully adopted to identify 
high-velocity components driven by powerful protostellar outflows.
\citep[e.g.,][]{stanke07,takahashi08,arce10,nakamura11a,nakamura11b}, 
and therefore we identify several molecular outflows in the L1641-N
cluster using the $^{12}$CO ($J=1-0$) data.

Molecular outflow activity in the L1641-N region was first found
by \citet{fukui86} at the position of the 
near-IR bright source IRAS 05338-0624 \citep[see also][]{fukui88}.
Further outflow surveys have been made through molecular line studies 
by several authors \citep[e.g.,][]{wilking90,stanke07}.
\citet{stanke07} conducted $^{12}$CO ($J=2-1$) mapping 
observations toward the L1641-N cluster using the IRAM 30 m
telescope and SMA interferometer, and identified a number of 
molecular outflow lobes.
Recently, \citet{davis09} carried out a near-infrared H$_2$ survey of 
outflows in Orion A including the L1641-N region.
To identify the molecular outflows, we first scrutinize the
velocity channel maps of our $^{12}$CO ($J=1-0$) 
(Figure \ref{fig:channel}) and the position-velocity diagrams 
to find localized blueshifted and redshifted emission.

In Figure \ref{fig:outflow}, we present the $^{12}$CO ($J=1-0$) integrated 
intensity contours overlaid on the 1.1 mm continuum image obtained 
by \citet{shimajiri10}.
The velocity intervals for the integration are $0 \sim 5$ 
km s$^{-1}$ and $10 \sim 16$ km s$^{-1}$ 
for blueshifted and redshifted components, respectively.
\citet{stanke07} labeled the outflow lobes identified from $^{12}$CO ($J=2-1$)
whose names are also designated in Figure \ref{fig:outflow}.
The most extended collimated redshifted lobe identified by
\citet{stanke07}, R-S, is not so prominent in our $^{12}$CO ($J=1-0$) map 
where this lobe is divided into several distinct knots. 
A number of H$_2$ knots labeled with SMZ 49 in \citet{davis09} 
are associated along this redshifted lobe.
The blueshifted emission elongated along the northeast to southwest line 
is also seen north of the dust peak near the cluster center. 
This corresponds to the B-NE and B-N lobes in \citet{stanke07}.
In the $^{12}$CO ($J=2-1$) map, this lobe has a Y shape, whereas
in our $^{12}$CO ($J=1-0$) map, the B-NE emission dominates over the faint 
B-N emission. This $^{12}$CO ($J=1-0$) blueshifted lobe also corresponds to
the one identified by \citet{fukui86}.
These are probably due to the fact that our $^{12}$CO ($J=1-0$) data tend to 
trace relatively low-velocity outflow components. In fact, 
according to \citet{stanke07}, at low velocities, the two lobes 
identified by the $^{12}$CO $(J=2-1)$ emission blend together with a broad 
patch of $^{12}$CO emission.  The most intense dust continuum emission 
is associated with the area between the redshifted (R-S) and 
blueshifted (B-NE and B-N) lobes,
which is in good agreement with the HCN and HCO$^+$ ridge 
found by \citet{fukui88} and the position of IRAS 05338-0624.
The H$_2$ flows labeled with SMZ 51 and SMZ 53 appear to follow 
the B-NE and B-N lobes, respectively \citep[see Fig. 6 of ][]{davis09}.

Another highly-collimated, redshifted lobe is seen to the west of the
R-S lobe, which
is in good agreement with the R-SW lobe in Figure 2 of \citet{stanke07}.
Similarly to the $^{12}$CO ($J=2-1$) map, this lobe appears not to have 
any obvious blueshifted counterlobe.
The R-W, B-E, and B-SE2 lobes identified by \citet{stanke07}
can be clearly recognized in our map.
The R-W lobe probably corresponds to the redshifted lobe 
identified by \citet{fukui86}.

From our $^{12}$CO map, the total outflow mass, momentum, and
energy are estimated to about 13 $M_\odot$, 80 $M_\odot$ km s$^{-1}$,
and 273 $M_\odot$ km$^2$ s$^{-2}$, respectively, 
by assuming the local thermodynamical equilibrium (LTE)
condition and optically-thin emission. 
Here, we adopt the same systemic velocity of 7.5 km s$^{-1}$ for 
all the outflow lobes and the values are corrected for
the inclination angle $\xi = 57.3^\circ$ \citep{bontemps96}.
The excitation temperature is adopted to 30 K, which is close to
the peak brightness temperature of the $^{12}$CO ($J=1-0$) line profile
at the L 1641-N cluster center.
We also integrated the physical quantities in the velocity 
intervals of $0 \sim 5$ km s$^{-1}$ and $10 \sim 16$ km s$^{-1}$ 
for the blueshifted and redshifted components, respectively.
We note that the physical quantities estimated above are 
insensitive to the assumed excitation temperature. 
The estimated quantities increase only by a few $\times 10$ \% over 
the range of $T_{\rm ex}= 20 - 50$ K.

Our $^{12}$CO map suggests that we identified the outflow lobes 
driven by about five embedded sources in the central part of L1641-N, 
the mean outflow momentum for a single
outflow may be estimated to be about  $80/5=16M_\odot$ km s$^{-1}$.
If we assume the median stellar mass of 0.5 $M_\odot$, then
this gives the outflow momentum per unit stellar mass of 
about 32 km s$^{-1}$, corresponding to the nondimensional 
parameter $f$ of about 0.32, which gauges the strength of an 
outflow, where the wind velocity of 100 km s$^{-1}$ is adopted.
This value is consistent with the fiducial values adopted by 
\citet{matzner00}, \citet{li06}, and \citet{nakamura07}.

\section{Discussion}
\label{sec:discussion}

\subsection{Triggered Formation of Dust Filaments}
\label{subsec:trigger}

As shown in Section \ref{sec:results}, some dust filaments are located 
near the cloud edge where the $^{12}$CO emission drops sharply, rather than
along the ridge of the parent filamentary cloud
(i.e., filaments A and E).
The dust continuum emission associated with the filaments also tend to 
have very steep gradients in the envelopes of the filaments, in
particular for the filament B.
In addition, the dust filaments have two-components with
different velocities.  The velocity difference between the two
components is significantly larger than the typical local turbulent
speed.
According to the line-width-size relation obtained by \citet{heyer04},
the typical turbulent velocity is estimated to be around 1 km s$^{-1}$, 
about 3 times smaller than the velocity difference observed 
in the dust filaments.
Thus, it is unlikely that the filaments were created by the 
dynamical compression due to the local turbulent flow.
From these observational facts, we here propose that the dust filaments 
in this region were created by external compression, instead of 
the spontaneous gravitational contraction.

\begin{figure}[h]
\plotone{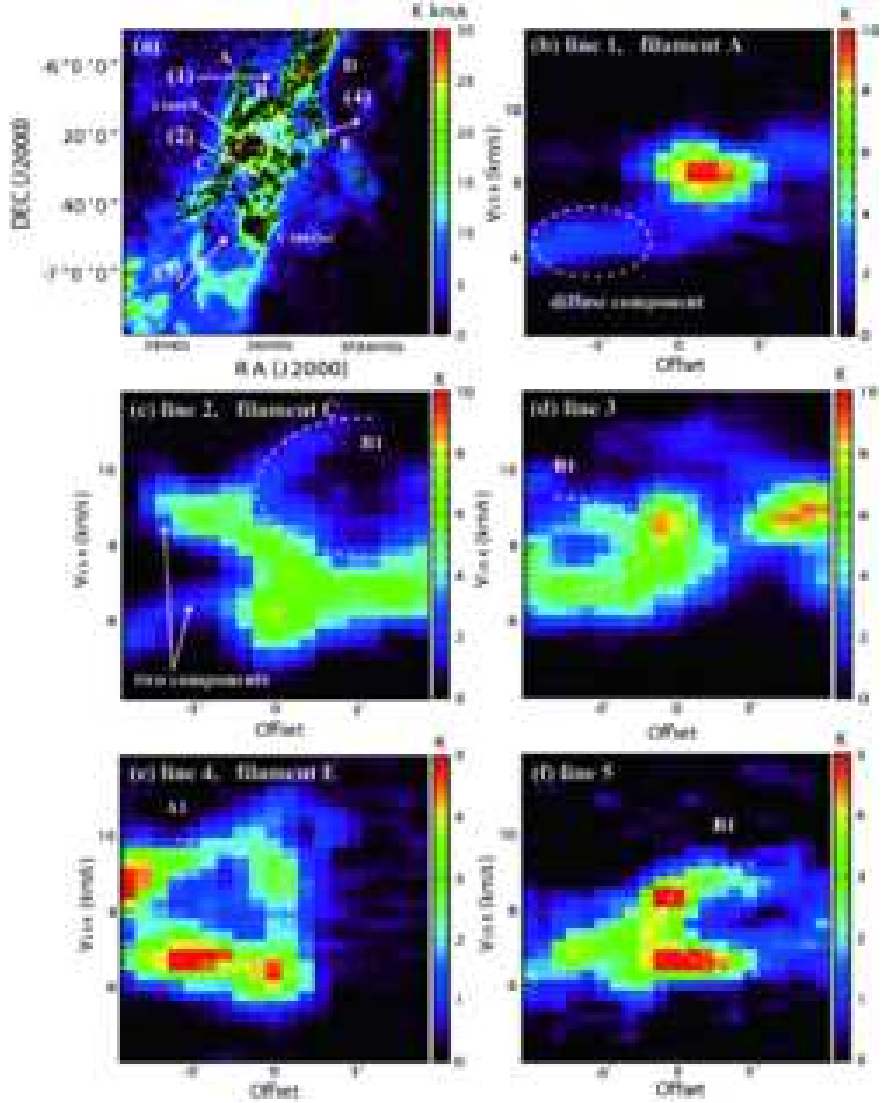}
\caption{(a) $^{13}$CO $(J=1-0)$ velocity integrated intensity map 
showing the positions of the PV diagrams. The velocity integration 
range is the same as that of \ref{fig:13co map}.
The lines (b) through (e) are the same as those of Figure \ref{fig:PV}.
The 1.1 mm continuum map is overlaid on the image with the 
black contours whose levels are the same as those of
 Figure \ref{fig:map1}.
Note that the southern part (below Dec. $\sim - 6:30$)
was not covered by the 1.1 mm observations \citep[see][]{shimajiri10}
(b) Position-velocity diagram along the  line (1).  
The velocity resolution of the data is 0.2 km s$^{-1}$.
(c) Same as panel (b) but for the line (2). 
(d) Same as panel (b) but for the line (3).
(e) Same as panel (b) but for the line (4).
(f) Same as panel (b) but for the line (5).}  
\label{fig:PV2}
\end{figure}

\begin{figure}[h]
\epsscale{0.8}
\plotone{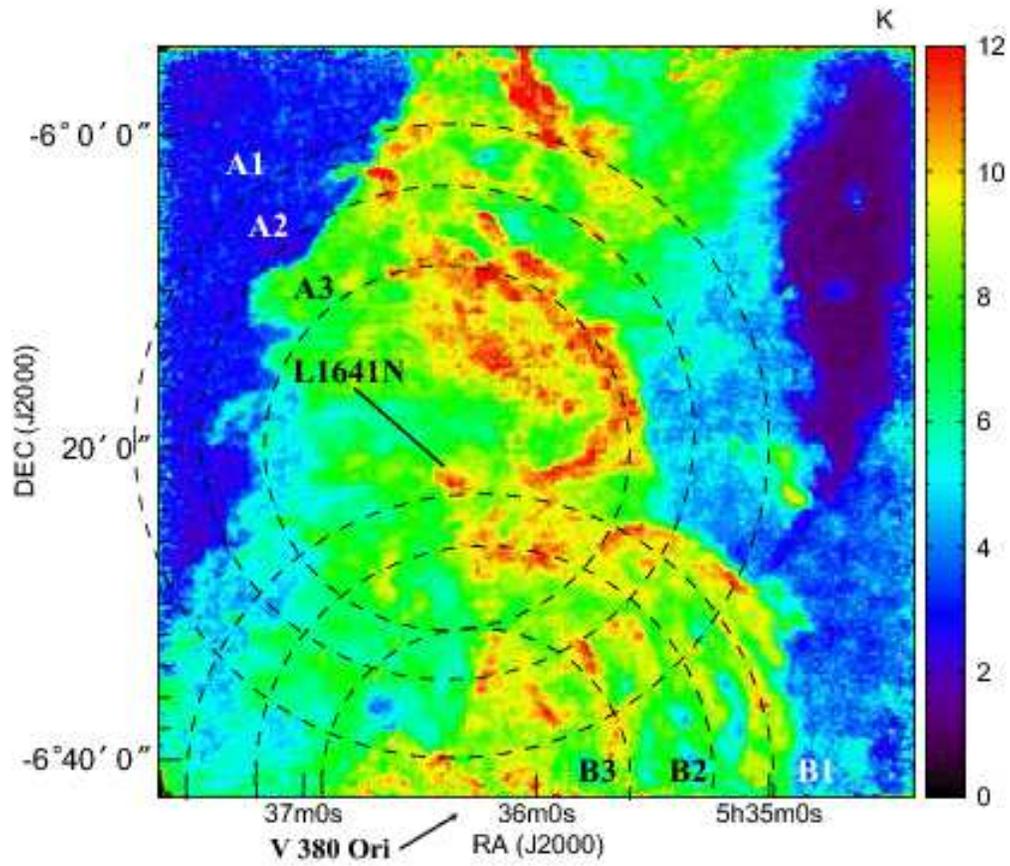}
\caption{Same as Figure \ref{fig:peak map} but 
the positions of parsec-scale shells are 
indicated with the dashed lines.
}  
\label{fig:peak map2}
\end{figure}

\begin{figure}[h]
\plotone{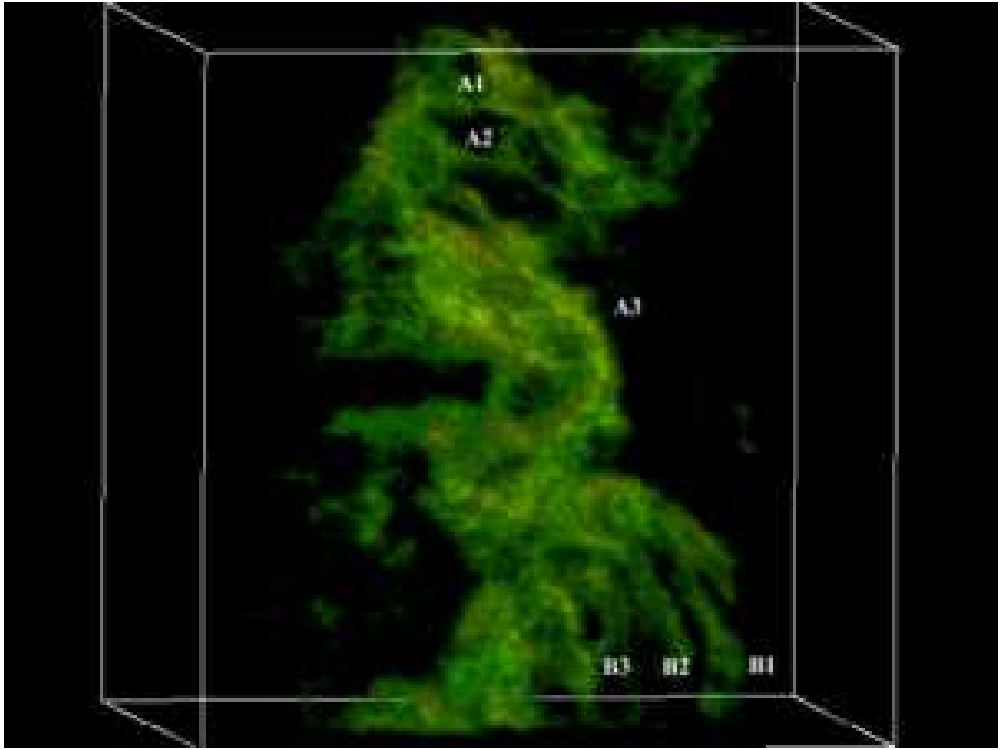}
\caption{Three-dimensional representation of the 
antenna temperature ($T_{\rm A}^*$) of the CO ($J=1-0$) emission
in the RA-DEC-$V_{\rm LSR}$ space.
The green color roughly represents the parts with
$T_{\rm A}^* \sim 15$ K.
Several arc-like structures can be recognized in the 
RA-DEC-$V_{\rm LSR}$ space.
}  
\label{fig:3d}
\end{figure}

\begin{figure}[h]
\plotone{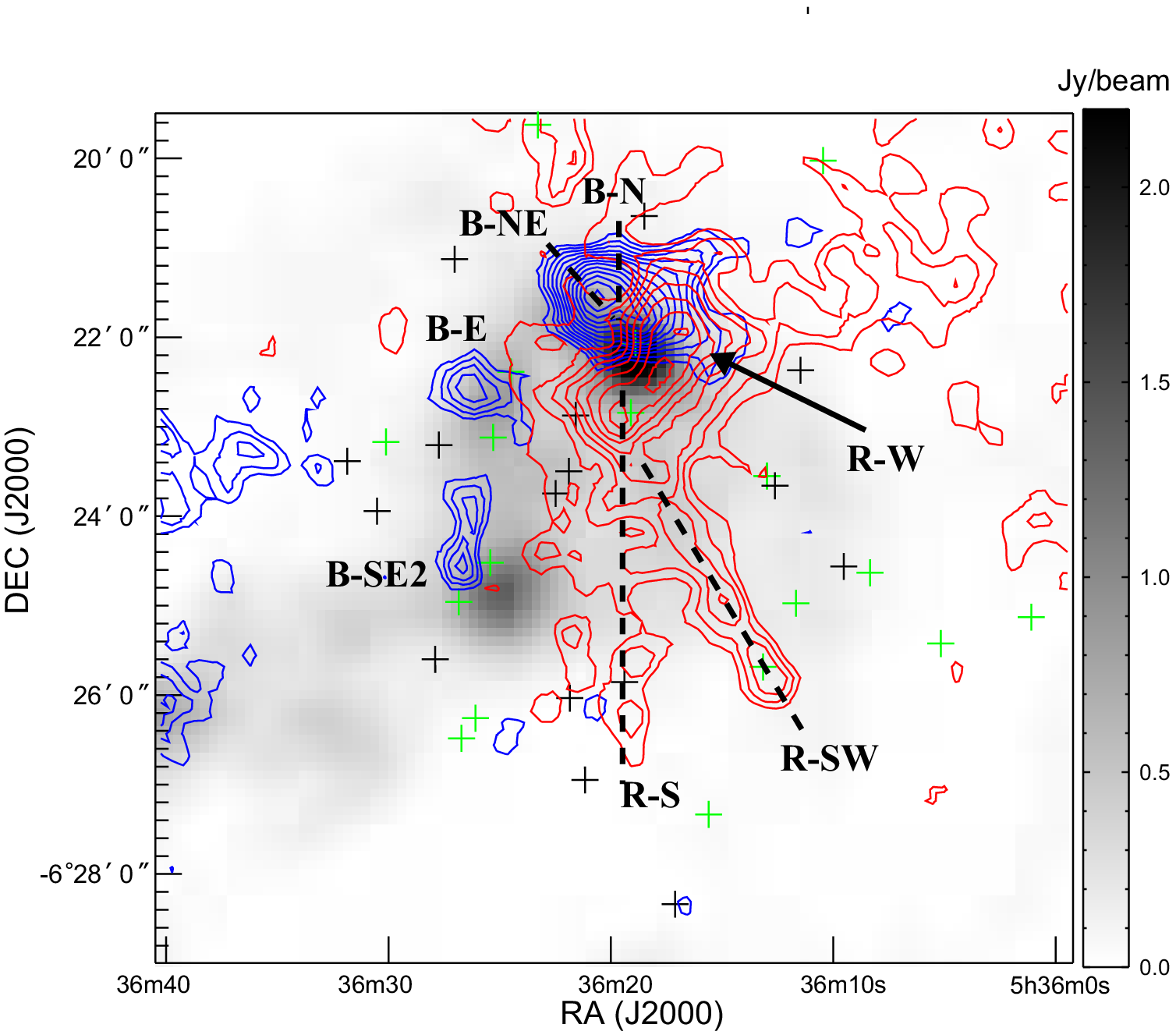}
\caption{
CO ($J=1-0$) integrated intensity contours toward 
the L1641-N cluster on the 1.1 mm image 
obtained by \citet{shimajiri10}.
The displayed area is almost the same as that shown in Figure 2
of \citet{stanke07} who mapped the same area in CO ($J=2-1$).
The outflow lobes seen in the CO ($J=1-0$) data are also labeled
following \citet{stanke07}.
The blue contours represent blueshifted CO ($J=1-0$) intensity
integrated from 0.25 km s$^{-1}$ to 5.25 km s$^{-1}$, starting from
5 K km s$^{-1}$ at intervals of 0.75 K km s$^{-1}$.
The red contours represent blueshifted CO ($J=1-0$) intensity
integrated from 9.75 km s$^{-1}$ to 15.75 km s$^{-1}$, starting from
9 K km s$^{-1}$ at intervals of 0.75 K km s$^{-1}$.
The gray scale represents the 1.1 mm image in units of Jy beam$^{-1}$.
The positions of the classical and weak-line T-Tauri stars 
observed by \citet{fang09} are shown in black and green crosses, 
respectively.
}  
\label{fig:outflow}
\end{figure}

On the larger scale of around 10 pc, 
there is a significant velocity gradient 
along the parent filamentary cloud.
The cloud component with velocities smaller than about 6 km s$^{-1}$ 
is dominant in the southern part 
of the parent filamentary cloud, whereas the component with 
velocities larger than about 7 km s$^{-1}$ is dominant in the northern
part (see Figure \ref{fig:3color} and Figure 2 of Bally et al. 1987).
The interaction between the two-components is likely to have 
created the filaments A, B, C, and D in this region.
Here, we refer to such a possible dynamical interaction as a  
cloud-cloud collision.  
Since the two components tend to overlap toward the dust filaments
on the plane of the sky 
(see Figures \ref{fig:PV}(b), \ref{fig:PV}(c), \ref{fig:PV2}(b), 
and \ref{fig:PV2}(c)), 
the collision may have occurred almost along the line of sight.
The collision may also have triggered star formation in the L1641-N
cluster as well as the formation of the massive filament C.
The cluster formation triggered by a cloud-cloud collision 
have been recently discussed by 
several authors for other star-forming regions 
\citep{xue08,furukawa09,duarte10,duarte11,galvan10, higuchi10}.

Besides the possible cloud-cloud collision, other parts of this region 
appear to have a velocity structure
that is in good agreement with an expanding motion created by
a shell.  A typical example is the filament E that has two 
velocity components inside the shell A1 but appears to 
converge into a single velocity component at the outer part 
(see Figures \ref{fig:PV}(e) and \ref{fig:PV2}(e)).
Such an arc-like structure in the position-velocity diagrams 
can be interpreted as an expanding motion of the shell.
The filament C, the most massive filament, also has 
such an arc-like structure in the position-velocity diagram
in the western side (Figure \ref{fig:PV2}(c)).
This massive filament may have recently been compressed by an 
expanding shell after it was created by the cloud-cloud collision
and the cluster formation was initiated.
The compression due to the expanding shell may have 
accelerated the recent star formation in the L1641-N cluster.
We don't think that the filaments A, B, C, and D were created by 
the expanding shells created by the stellar feedback from
the protoclusters because their main axes tend to cross 
the shells and they appear more or less straight, 
rather than curved like the filament E.

Hence, the cloud-cloud collision and the expanding shells 
appear to have influenced the cloud structure and kinematics 
in this region.
In Sections \ref{subsec:cloud-cloud collision} and \ref{subsec:shell},
 we discuss these two dynamical events, 
the cloud-cloud collision and the expanding shell, 
in detail.

\subsection{Cloud-Cloud Collision}
\label{subsec:cloud-cloud collision}

According to our CO channel maps,  the two-components with velocities 
of $4-6$ km s$^{-1}$ and $7-12$ km s$^{-1}$ appear to be 
interacting to form the dense filaments.
The interpretation is consistent with the previous $^{13}$CO 
($J=1-0$) observations which suggested that two or more different 
velocity components are associated with the Orion A giant molecular 
cloud complex \citep{bally87,bally89}.
The velocity difference between the two-components is 
about 3 km s$^{-1}$, corresponding to the Mach number of 
10 under the assumption that the gas temperature is 20 K.
The supersonic collision between the two-components 
would have increased the local density by a factor of $10^2$
if the shock is isothermal. 
Therefore, such a shock
can create the dense filaments whose densities are 
as large as $10^{4-5}$ cm$^{-3}$ from preshock gas with 
$10^{2-3}$ cm$^{-3}$.
In addition, the shock crossing time is estimated to be
a few $\times 10^5 - 10^6$  yr when the typical size of 
the filaments is adopted as a few pc.
The estimated shock crossing time is  shorter than 
the lifetime of the L1641-N cluster (a few $\times 10^6$ yr,
see, e.g., \citet{hodapp93} and \citet{galfalk08})
by a factor of a few.  Consequently, it is possible for 
such a collision to have triggered star formation in the dense filaments
of this area.
Since the shock crossing time is somewhat shorter than the lifetime of 
the cluster, the filament is expected to be in the postshock stage 
in which the observed two components may be passing over, thus 
leaving away from each other. 
According to \citet{sakamoto97}, the blueshifted component is 
extended over the larger low-density area toward the eastern part 
of the cloud. This fact suggests that the collision with large clouds
or flows (1$-$10 pc), instead of the collision with smaller clouds, 
may have occurred.
The dynamical interaction with such external flow 
has been recently suggested by \citet{shimajiri10} for 
the northern area of the Orion A giant molecular cloud complex.
The diffuse blueshifted component found from our $^{12}$CO data
may be related to the flow pointed out by \citet{shimajiri10}.

Recently, an interesting scenario on the formation of the Orion A
giant molecular cloud complex 
has been proposed by \citet{hartmann07} who performed 
two-dimensional numerical simulations of gravitational collapse 
of a thin gas sheet. According to their model, two large filaments 
are first created near the edge of the gas sheet by the gravitational 
effects of the cloud edge where
the gravitational potential takes its local minimum
(see also Larson 1976 and Bastien 1983 for the effect of the cloud edge 
on the gravitational fragmentation). 
They demonstrated how an elongated rotating gas sheet with a density gradient
along the major axis can gravitationally collapse to produce 
a structure qualitatively resembling the whole Orion A giant molecular
cloud complex,
having a fan-shaped structure at the southern part, ridges along the
fan, and a narrow integral shaped filament at the northern part.
In their model, our observed area, L1641-N, is located at the 
intersection between the fan and the narrow main filament, where
the two large filaments collide with each other. 
Such a large scale interaction might be able to explain the origin of 
the dust filaments, although it remains unclear how the hypothetical 
parent gas sheet was created.

\subsection{Parsec-Scale CO Shells and Protocluster Winds}
\label{subsec:shell}


As shown in Figure \ref{fig:peak map2}, our CO map revealed that several 
shells appear to shape the cloud structure in this region.  
The radii of the circles are about 10$'-$20$'$, corresponding to about 
1$-$2 pc at a distance of 400 pc.
Interestingly, the centers of the circles A1, A2, and A3 are 
close to the L1641-N cluster center, and those of the circles 
B1, B2, and B3 to the V 380 Ori center.
Such large shells are unlikely to form by stellar feedback from 
a single young star unless it is massive.  Since the spectral types 
of the most massive cluster members are late B or early A for the
two clusters, the shells may be difficult to create only by  
the stellar feedback of the most massive stars.   
Therefore, we suggest that these parsec-scale shells  
were created by multiple winds and/or outflows 
from cluster member YSOs.
The reason why three or more shells are associated with each cluster remains
unclear. One possibility is that the total momentum injection rate 
from cluster member YSOs is not constant but episodic 
because the star formation rate is likely to fluctuate 
or oscillate with time.
Another possibility is that these circles represent dense parts of 
an expanding shell that is propagating into inhomogenius media
Here, we call these YSO winds as ``protocluster wind.''
In the next subsection, we consider the dynamical evolution
of the shell driven by the protocluster wind, using a simple 
analytic model.

Kinematical evidence that the protocluster wind is responsible for 
the cloud dynamical evolution comes from 
the filament E that is located at the western edge of the cloud and 
has an arc-like shape. 
The filament E has two velocity components inside the 
circle A1 but appears to converge into a single velocity 
component at the outer part (see Figures \ref{fig:PV}(e) and
\ref{fig:PV2}(e)), as mentioned in Section \ref{subsec:trigger}.
It is also distributed along the circle A1 whose center coincides with 
the L1641-N cluster, suggesting that it may have been created by the 
 shell driven by the protocluster wind from the L1641-N cluster.

\begin{figure}[h]
\epsscale{0.6}
\plotone{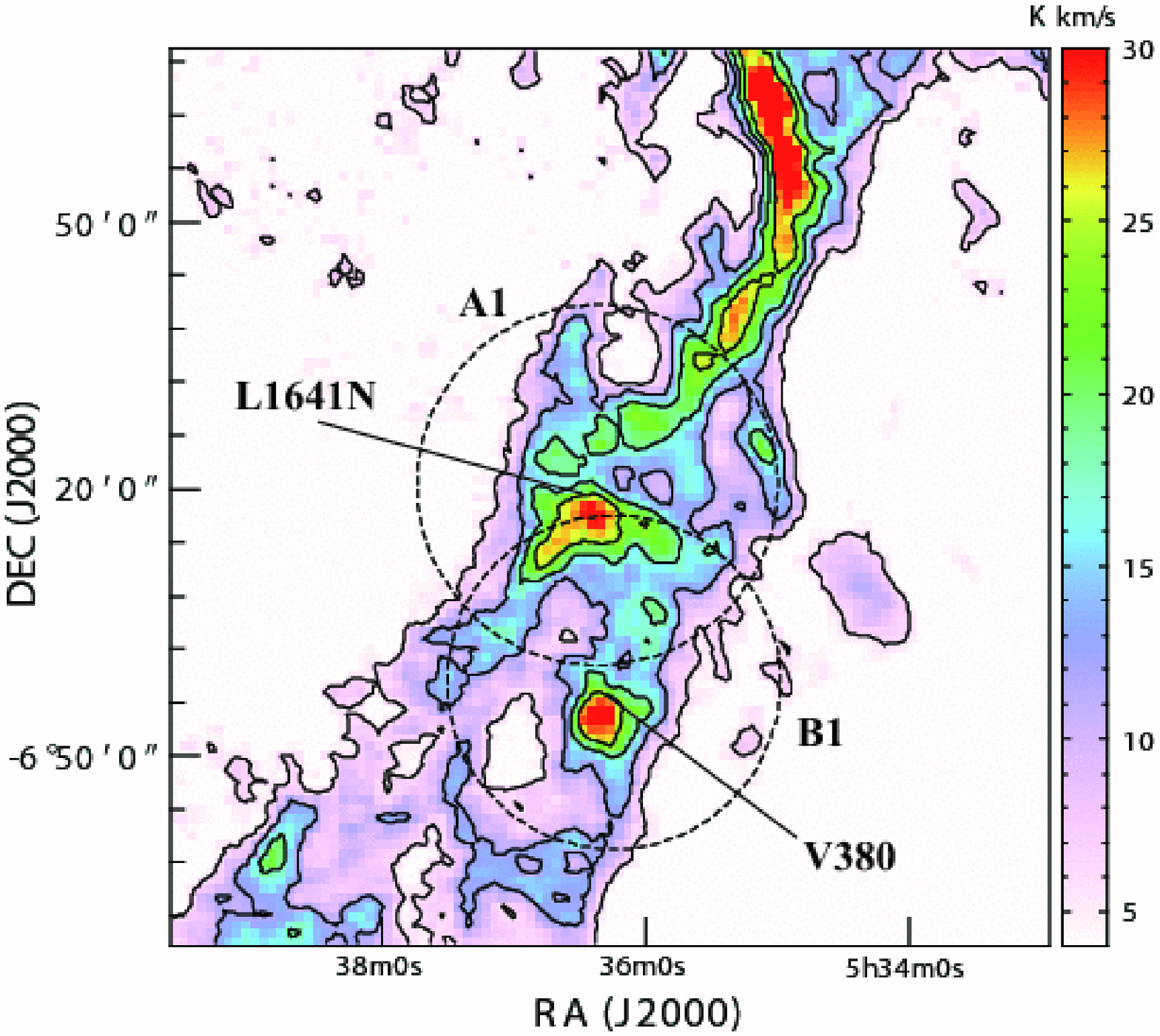}
\plotone{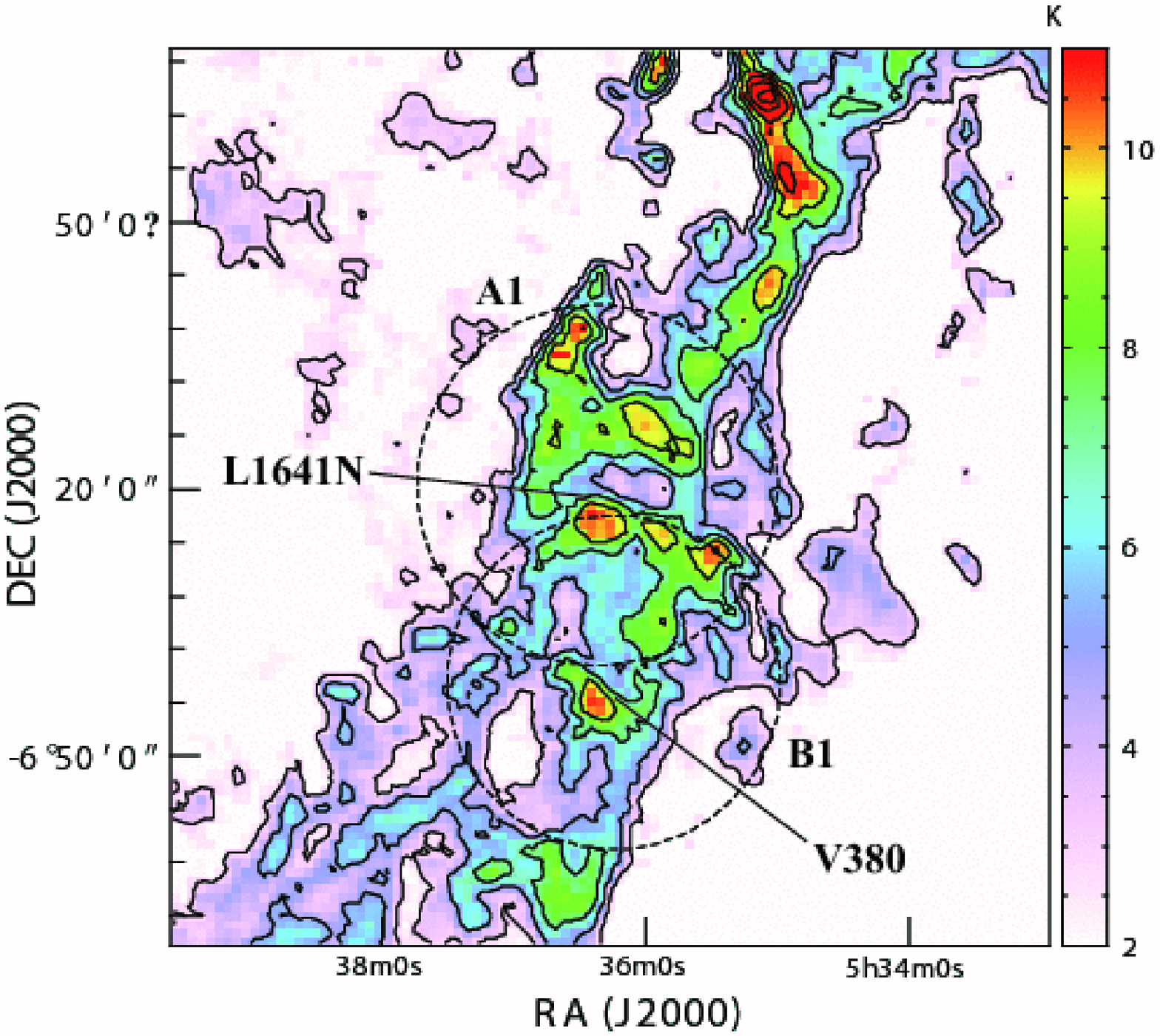}
\caption{(a) $^{13}$CO total integrated intensity map
in the range from $V_{\rm LSR}=1$ km s$^{-1}$ to
14 km s$^{-1}$ \citep{bally87}.
(b) $^{13}$CO peak intensity map in the same area presented 
in panel (a).
For both the panels, the positions of the circles A1 and B1 presented 
in Figure \ref{fig:peak map2} are indicated with the dashed lines.
}  
\label{fig:13co+shell}
\end{figure}

Another example of the parsec-scale expanding shell is the
one whose center is close to V 380 Ori. This shell is fully covered 
by the $^{13}$CO ($J=1-0$) map obtained by \citet{bally87}, although
the southern part of the shell is not covered by our $^{12}$CO data.
We show in the $^{13}$CO integrated intensity map presented in 
Figure \ref{fig:13co+shell} the circles that appear to fit the shells.  
The $^{13}$CO position-velocity diagrams of the shells toward the two 
positions (Figures \ref{fig:PV2}(d) and \ref{fig:PV2}(f)) clearly 
show the existence of the two different velocity components, 
consistent with the expanding motion.
The expanding motion of the shell is in good agreement with
\citet{heyer92} who found a similar expanding motion of the shell.

\subsection{Expanding Motions Driven by Protocluster Winds}

In the following, we discuss how the protocluster wind evolves in
the parent molecular cloud using a simple analytic model.
Here, we consider the motion of an expanding shell driven by
protostellar winds from cluster member YSOs in a uniform 
media with density $\rho_0$.
For simplicity, we don't take into account the effects of 
protostellar outflows in the following calculations, but 
discuss the effect of the protostellar outflow 
feedback in the next subsection.

If a wind from a protostar with a radius $R_*$ 
is driven by a ram pressure $\rho_{\rm w} V_{\rm w}^2$, then
the motion of an expanding shell is described as 
\begin{equation}
\frac{d}{dt}\left(\frac{4\pi}{3}R^3 \rho_0 \frac{dR}{dt}\right)
=\sum _{i=1}^N 4\pi R_*^2 \rho_{\rm w} V_{\rm w}^2
\end{equation}
where $\rho_{\rm w}$ is the wind density, $V_{\rm w}$ is the wind
velocity, $R$ is the radius of the shell, and $N$ is the number of YSOs.
For simplicity, all the stars are assumed to inject the wind momentum
simultaneously at the same constant rate.  
From the above equation, the time evolution of the shell radius
is given by
\begin{eqnarray}
R &=& \left(\frac{3\sum_{i=1}^{N}\dot{M} V_{\rm w}}{2\pi \rho_0}\right)^{1/4}
 t^{1/2} \nonumber \\
&\simeq& 2 \ {\rm pc}\ \left(\frac{N}{100}\right)^{1/4}
\left(\frac{\dot{M}}{10^{-7} M_\odot {\rm yr}^{-1}}\right)^{1/4}
\left(\frac{V_{\rm w}}{200 \ {\rm km \ s}^{-1}}\right)^{1/4}  \nonumber \\
&& \times \left(\frac{\rho_0}{10^3 \ {\rm cm}^{-3}}\right)^{-1/4}
\left(\frac{t}{10^6 \ {\rm yr}}\right)^{1/2} \ ,
\label{eq:radius}
\end{eqnarray}
where $\dot{M}$ is the mass loss rate from a single YSO 
and given by $4\pi R_*^2 \rho_{\rm w} V_{\rm w}$.
The expanding velocity of the shell, $V_s$, is evaluated as
\begin{eqnarray}
V_s&=&\frac{dR}{dt}  \nonumber \\
&\simeq& 1 \ {\rm km \ s}^{-1}\ \left(\frac{N}{100}\right)^{1/4}
\left(\frac{\dot{M}}{10^{-7} M_\odot {\rm yr}^{-1}}\right)^{1/4}
\left(\frac{V_{\rm w}}{200 \ {\rm km \ s}^{-1}}\right)^{1/4}  \nonumber \\
&& \times \left(\frac{\rho_0}{10^3 \ {\rm cm}^{-3}}\right)^{-1/4}
\left(\frac{t}{10^6 \ {\rm yr}}\right)^{-1/2} \ .
\label{eq:velocity}
\end{eqnarray}
The typical mass loss rate and wind velocity from YSOs are somewhat uncertain.
Some observations suggest that the mass loss rate and wind velocity 
from a low-mass YSO are of the order of $10^{-7} M_\odot$ yr$^{-1}$ and 
a few $\times 10^2$ km s$^{-1}$, respectively 
\citep[e.g.,][]{norman80,wilkin98}. Here, we adopted 
$\dot{M}=10^{-7} M_\odot$ yr$^{-1}$ and 
$V_{\rm w}=200$ km s$^{-1}$ as their representative values.
For the L1641-N and V 380 Ori clusters, the number of cluster member
YSOs is of the order of 100.  
Adopting the density of $10^3$ cm$^{-3}$ for the ambient gas and 
the cluster lifetimes of $10^6$ yr, the radius of the expanding shell 
and the expansion velocity are evaluated to be about 2 pc
and 1 km s$^{-1}$, respectively, comparable to those of the shells 
identified from our $^{12}$CO data.
Therefore, the protocluster winds from these 
two clusters are likely to have enough energies to produce the 
parsec-scale shells within a few Myr.

\subsection{Molecular Outflows in the L1641N cluster}

In the previous subsection, we omitted the contribution of 
the outflow feedback on the dynamical evolution of the expanding shell.
However, recent numerical simulations of cluster formation have 
suggested that protostellar outflow feedback plays an important
role in regulating star formation in cluster-forming clumps
because they inject substantial amount of kinetic energy into
the surroundings \citep{li06,nakamura07,carroll09,wang10}.
In addition, the propagation directions of the collimated outflows 
in the clustered environment are not so aligned preferentially
in the global magnetic field direction even in
the presence of a strong magnetic field \citep{nakamura11}.
Therefore, the total outflow momentum is expected to
be injected isotropically on average.
Here, we attempt to assess how the outflow feedback 
of the L1641-N cluster influences the surrounding gas
and contributes to the dynamical evolution of the expanding shell.

\subsubsection{Dynamical State of the L1641-N Cluster Forming Clump}

Recently, \citet{nakamura11b} investigated how molecular outflow
feedback influences the dynamical state of a nearby cluster-forming
clump, Serpens South, by applying the virial analysis.
They found that the Serpens South clump is close to virial equilibrium
and the total kinetic energy injected by the current outflow activity
is less than the clump gravitational energy, concluding that
the current outflow activity is not enough to destroy the whole 
cluster-forming clump.
In the following, we follow \citet{nakamura11b} and clarify
how the current outflow activity in the L1641-N cluster
impacts the whole clump.

The virial equation for a spherical clump is given by
\begin{equation}
\frac{1}{2}{\partial ^2 I \over \partial t^2} = 2U + W \ ,
\label{eq:virial}
\end{equation}
where the terms, $I$, $U$, and $W$, denote the moment of inertia,
internal kinetic energy, and gravitational energy, respectively.
Here, we neglect the surface pressure term.
A clump is in virial equilibrium when the right-hand-side of 
Equation (\ref{eq:virial}) is zero, i.e., $2U+W=0$.
The kinetic energy and gravitational energy terms are expressed, 
respectively, as 
\begin{equation}
U={3M\Delta V^2 \over 16 \ln 2}
\end{equation}
and 
\begin{equation}
W=-a {GM^2 \over R} \left[1-\left({\Phi \over \Phi_{\rm
                             cr}}\right)^2\right] \ ,
\end{equation}
where $M$ is the clump mass, $\Delta V$ is the 1D FWHM velocity width,
$G$ is the gravitational constant, $R$ is the radius of the clump, 
the values $\Phi$ and $\Phi_{\rm cr}$ are, respectively,
the magnetic flux penetrating the clump and the critical magnetic flux
above which the magnetic field can support the clump against 
the self-gravity. 
Although according to recent observations of magnetic fields
associated with the nearby parsec-scale cluster-forming clumps,
the values of $\Phi / \Phi_{\rm cr}$ are estimated to be
around 0.5 \citep[e.g.,][]{falgarone08,sugitani10,sugitani11,kwon11}, 
we here neglect the effect of the magnetic support for simplicity
(i.e., $\Phi=0$). 
The dimensionless parameter, $a$, 
measures the effects of a nonuniform and/or nonspherical mass
distribution \citep{bertoldi92}, and is of the order unity.
For a uniform sphere and a centrally-condensed sphere with
$\rho \propto r^{-2}$, $a=3/5$ and 1, respectively.
Here, we adopt $a=1$ because the cluster-forming clump tends
to be centrally condensed.

According to \citet{stanke07}, the radius and mass of the 
cluster-forming clump
associated with the L1641-N cluster are roughly estimated to be
0.5 pc and 150 $M_\odot$, respectively. Here, 
only the part involved in directly building the L1641-N cluster, 
i.e., a head of the clump, is defined as the cluster-forming clump,
although the parent clump has a long tail extending southeast 
of the L1641-N cluster, as shown in the 1.1 mm map.
The 1D FWHM velocity width is estimated at about 2 km s$^{-1}$
from the $^{12}$CO data.
Using these values, the kinetic energy and gravitational energy terms
are estimated to be $2U\simeq 325 M_\odot$ km$^2$ s$^{-2}$ and 
$W\simeq -194 M_\odot$ km$^2$ s$^{-2}$, respectively.
These values are comparable to the estimate by \citet{stanke07} who
conducted a similar analysis using their $^{12}$CO ($J=2-1$) data.
Thus, the clump is likely to be close to virial equilibrium
or somewhat gravitationally unbound.
On the other hand, the total kinetic energy due to the outflows is 
evaluated at $E_{\rm out} \simeq 273 M_\odot$ km$^2$ s$^{-1}$.
This is comparable to the clump gravitational energy.  
Although the substantial amount of the outflow kinetic energy 
appears to escape from the clump, the energy 
input due to the current outflow activity seems to influence 
the clump dynamics significantly.

\subsubsection{Effects of the Molecular Outflows on the 
Expanding Shells}

As shown in Figure \ref{fig:outflow}, the identified outflow 
lobes often extend beyond the dense clump traced by the dust 
continuum emission, implying that the outflow momentum escaped 
from the clump is likely to influence 
the density and velocity structure in the surroundings. 
If the escaped momentum fllux is about a half the total injected outflow 
momentum flux, it is estimated to be
about $10^{-3} M_\odot$ km s$^{-1}$ yr$^{-1}$,
assuming the outflow dynamical time of a few $\times 10^4$ yr.
On the other hand, the total momentum flux injected from 
the protostellar winds is estimated to be 
$N \dot{M} V_{\rm w} \simeq 100 \times 10^{-7} \times 200 = 
2\times 10^{-3} M_\odot$ km s$^{-1}$ yr$^{-1}$, where we adopted 
the values given in Equations (\ref{eq:radius})
and (\ref{eq:velocity}).
The estimated wind momentum flux is comparable to that of 
the protostellar outflow feedback.
Therefore, we conclude that both the protostellar outflow and
wind feedback can contribute to the dynamical evolution of the expanding
shell created by the protocluster wind from the L1641-N cluster.
However, the weak dependence on the injected momentum in 
Equations (\ref{eq:radius}) and (\ref{eq:velocity})
indicates that even if we take into account both the protostellar 
outflows and winds, the estimated radius and expanding velocity 
of the shell do not change significantly.

\section{Summary}
\label{sec:summary}

We carried out the $^{12}$CO ($J=1-0$) mapping observations toward 
the $48'\times 48'$ area including the L1641-N cluster in
the Orion A giant molecular cloud complex, using the Nobeyama 45 m telescope.
The main results are summarized as follows.

1. From comparison between the $^{12}$CO ($J=1-0$) map and the 1.1 mm continuum
image, we found that several dust filaments are located near the cloud edge 
traced by the $^{13}$CO ($J=1-0$) emission.

2. The dust filaments have two-components with different
velocities.  The velocity difference between the two-components
is about 3 km s$^{-1}$, which is significantly larger than the
typical local turbulent speed of 1 km s$^{-1}$. 
Therefore, we suggest that the dust filaments were 
created by the dynamical compression triggered externally, 
i.e., a cloud-cloud collision, 
instead of the spontaneous gravitational contraction.

3. The $^{12}$CO ($J=1-0$) and $^{13}$CO ($J=1-0$) velocity channel maps
suggest that the blueshifted ($V_{\rm LSR}\lesssim 6$ km s$^{-1}$) 
and redshifted ($V_{\rm LSR}\gtrsim 7$ km s$^{-1}$) components are
interacting with each other. Since the two components appear to overlap
toward the dust filaments 
on the plane of the sky, the collision between the two components 
may have occurred almost along the line of sight.
A good example of such a cloud-cloud collision 
is the most massive dust filament, 
associated with the L1641-N cluster, which has the two-components 
with different velocities at the long tail stretching from the 
head of the dust filament.  
We suggest that the formation of the L1641-N cluster may have been 
triggered by such a collision.
Since the shock crossing time is somewhat shorter than the lifetime 
of the L1641-N cluster, the massive filament associated with the cluster
may be in the postshock stage in which the two components with 
the different velocities are passing over, thus leaving away from each other.

4. We found several parsec-scale shells in the $^{12}$CO ($J=1-0$) data cube.  
Some of the shells appear to be spatially well-ordered and homocentric.
The centers of the shells are close to either the L 1641-N or 
V 380 Ori cluster centers, implying that the star formation 
activity in the clusters may be responsible for 
the formation and evolution of the shells.
In particular, the shell surrounding V 380 Ori is prominent
in the $^{13}$CO map.

5. The molecular gas distribution and kinematical structure 
of this region led us to the following scenario.
On the large scale of at least about 1 $-$ 10 pc, 
a cloud-cloud collision may have occurred almost along 
the line of sight in this region, contributing to the 
formation of several dense filaments (the filaments A through D).  
The cloud-cloud collision have triggered the formation of the L1641-N cluster.
Multiple protostellar winds and outflows from the 
cluster member YSOs created large expanding bubbles
that can be recognized in the $^{12}$CO and $^{13}$CO maps.
Here, we call these YSO winds as ``protocluster winds.''
The dust filament located at the western cloud edge (filament E) appears
to curve along the shell whose center is the L1641-N cluster,
and have two different velocity components at the inner part
of the shell. The two components appear to converge into 
a single velocity component at the outer part of the shell.
It is consistent with an idea that the filament E is 
a part of the expanding shell created by the protocluster wind 
from the L1641-N cluster. 
The shell surrounding V 380 Ori also has two different 
velocity components both in the $^{12}$CO and $^{13}$CO maps,
reaching the filament C.
It presumably hit the filament C, influencing the recent star 
formation in the L1641-N cluster.
Both the cloud-cloud collision and the protocluster winds 
are likely to have created the complicated cloud morphology and 
kinematics in this region.

6. Toward the L1641-N cluster, we identified a number of the outflow 
lobes using the $^{12}$CO ($J=1-0$) data.  The identified lobes are 
reasonably in good agreement with the results of \citet{stanke07}
who identified the outflow lobes in this region with the $^{12}$CO ($J=2-1$)
emission, using the IRAM 30 m telescope and 
SMA. 
The total outflow energy in the L1641-N cluster 
is comparable to the gravitational energy of the cluster-forming clump. 
This may suggest that the outflow feedback influences the dynamical 
evolution of the clump significantly. 
Assuming the median stellar mass of 0.5$M_\odot$, 
the mean outflow momentum per unit stellar mass is estimated to be 
about 32 km s$^{-1}$, under the assumption of optically thin gas. 
This mean outflow momentum corresponds to the nondimensional 
outflow parameter of $f=0.32$, which gauges the strength of an outflow.
This value of $f$ is comparable to  the fiducial values
of $f = 0.4$ adopted by \citet{matzner00}, \citet{li06}, and 
\citet{nakamura07}.

\acknowledgments 
This work is supported in part by a Grant-in-Aid for Scientific Research
of Japan (20540228, 22340040). 
We thank John Bally for kindly giving us the $^{13}$CO ($J=1-0$) 
fits data of the Orion A giant 
molecular cloud complex taken by the 7 m telescope of AT\&T Bell Laboratories.
We are grateful to Henrik Beuther, Christopher J. Davis, 
M. S. Nanda Kumar, and Christopher F. McKee for their valuable comments.
This work was carried out as one of the projects of the Nobeyama 
Radio Observatory (NRO), which is a branch of the National Astronomical 
Observatory of Japan, National Institute of Natural Sciences. 
We also thank the NRO staff for both operating the 45 m and helping us
with the data reduction.

\end{document}